\documentclass[twocolumn, twocolappendix,numberedappendix]{openjournal}

\graphicspath{{./}{figures/}}
\renewcommand{\baselinestretch}{1.075}
\newcommand\numberthis{\addtocounter{equation}{1}\tag{\theequation}}
\shortauthors{}
\usepackage{macros}
\usepackage{natbib, booktabs}
\usepackage[backref,breaklinks,colorlinks]{hyperref}
\hypersetup{linkcolor=blue,citecolor=blue,filecolor=blue,urlcolor=blue}
\usepackage{orcidlink}

\defcitealias{schutt_dark_2025}{S25}
\defcitealias{zhang_general_2023}{Z23}
\newcommand{\schutt}{\citetalias{schutt_dark_2025}\xspace}
\newcommand{\zhang}{\citetalias{zhang_general_2023}\xspace}

\newcommand{\gtwo}{e^{(2)}}
\newcommand{\ttwo}{T^{(2)}}
\newcommand{\efour}{e^{(4)}}
\newcommand{\tfour}{T^{(4)}}

\newcommand{\vonkarman}{von K\'arm\'an\xspace}
\newcommand{\psfws}{\texttt{psf-weather-station}\xspace}
\newcommand{\galsim}{\texttt{GalSim}\xspace}
\newcommand{\lsstpipe}{LSST Science Pipelines\xspace}
\newcommand{\piff}{\texttt{PIFF}\xspace}
\newcommand{\ngmix}{\texttt{ngmix}\xspace}
\newcommand{\imsim}{\texttt{imSim}\xspace}
\newcommand{\montauk}{\texttt{montauk}\xspace}
\newcommand{\batoid}{\texttt{batoid}\xspace}
\newcommand{\descwlshearsims}{\texttt{descwl-shear-sims}\xspace}
\newcommand{\descwlpackage}{\texttt{WeakLensingDeblending}\xspace}
\newcommand{\opsim}{\texttt{OpSim}\xspace}
\newcommand{\metacal}{\texttt{metacalibration}\xspace}
\newcommand{\metadet}{\texttt{metadetection}\xspace}
\begin{document}

\title{Simulation tests of PSF modeling for cosmic shear with the Vera C. Rubin Observatory}

\author{
Claire-Alice H\'ebert\orcidlink{0000-0002-7397-2690}$^{1\ast}$}
\author{Erin S. Sheldon$^{1}$}
\author{Tianqing Zhang\orcidlink{0000-0002-5596-198X}$^{2}$}
\author{Joachim Harnois-D\'eraps\orcidlink{0000-0002-4864-1240}$^{3}$}
\author{Mike Jarvis$^{4}$}
\author{the LSST Dark Energy Science Collaboration}

\affiliation{$^{1}$Physics Department, Brookhaven National Laboratory, Upton, NY 11973, USA}
\affiliation{
$^{2}$Department of Physics and Astronomy and PITT PACC, University of Pittsburgh, Pittsburgh, PA 15260, USA}
\affiliation{$^{3}$School of Mathematics, Statistics and Physics, Newcastle University, Herschel Building, NE1 7RU, Newcastle-upon-Tyne, UK}
\affiliation{$^{4}$Department of Physics and Astronomy, The University of Pennsylvania, Philadelphia, PA 19104, USA}
% \author{
% Claire-Alice H\'ebert\orcidlink{0000-0002-7397-2690}$^{1\ast}$,
% Erin S. Sheldon$^{1}$,
% Tianqing Zhang\orcidlink{0000-0002-5596-198X}$^{2}$,
% Joachim Harnois-D\'eraps\orcidlink{0000-0002-4864-1240}$^{3}$,
% and Mike Jarvis$^{4}$
% }
% \affiliation{
% $^{1}$Physics Department, Brookhaven National Laboratory, Upton, NY 11973, USA\\
% $^{2}$Department of Physics and Astronomy and PITT PACC, University of Pittsburgh, Pittsburgh, PA 15260, USA\\
% $^{3}$School of Mathematics, Statistics and Physics, Newcastle University, Herschel Building, NE1 7RU, Newcastle-upon-Tyne, UK\\
% $^{4}$Department of Physics and Astronomy, The University of Pennsylvania, Philadelphia, PA
% 19104, USA
% }
\email[$^\ast$ email: ]{chebert38@alumni.stanford.edu}

\begin{abstract}
Exceptional control of systematic effects is required in order to achieve unbiased cosmic shear two-point correlation function measurements with the next generation of galaxy imaging surveys, such as the Vera C. Rubin Observatory Legacy Survey of Space and Time (LSST). One critical challenge is accurately modeling the point-spread function (PSF), as errors in PSF estimation can introduce spatially correlated biases in galaxy shape measurements. The LSST Science Pipelines, which will be used to process Rubin data, include an implementation of the \piff (PSFs in the Full Field of View) package originally developed for, and demonstrated to perform well on, DES-Y3 and Y6 data. In this work we use semi-realistic image simulations, mimicking LSST observing conditions in $i$-band over 100 square degrees, to perform an end-to-end test of the Rubin PSF modeling pipeline. PSF model residuals are quantified using both second- and fourth-order moment parameters and performance is evaluated, for both LSST year 1 and year 10 depth, with a series of diagnostic tests. We find that the additive bias contribution from PSF modeling errors to the non-tomographic cosmic shear data vector is well below 30\% of the cosmic shear uncertainty estimated from our analytic covariance matrix. Though our simulations exclude several known effects that may further challenge PSF modeling, these results demonstrate promising performance from \piff on LSST-like data and provide an early benchmark for ongoing PSF validation efforts for LSST weak lensing analyses. 
\end{abstract}

\section{Introduction}\label{sec:intro}
Cosmic shear refers to the weak lensing of light from distant galaxies by the
gravitational potential of the large-scale distribution of matter in the universe.
These lensing distortions, which cause the images of galaxies to appear slightly
sheared, are thus sensitive to the matter distribution along the line of sight.
Cosmic shear, which is most often measured via the two-point correlation function
of observed galaxy shapes, has become one of the most constraining probes of
the dark energy equation of state parameters \citep{albrecht_report_2006, kilbinger_cosmology_2015, mandelbaum_weak_2018}.

High-precision cosmic shear results depend on accurate measurements of very small ($\sim$\,2$\%$) shears, and large catalogs of galaxies. Several ``stage-III'' imaging
survey collaborations have successfully performed these measurements to date: the
Kilo-Degree Survey \citep[KiDS,][]{asgari_kids-1000_2021, wright_kids-legacy_2025},
the Dark Energy Survey \citep[DES,][]{troxel_dark_2018, secco_dark_2022, amon_dark_2022},
and the Hyper-Suprime Cam Strategic Survey Program \citep[DES,][]{hamana_cosmological_2020, li_hyper_2023}.
The Legacy Survey of Space and Time \citep[LSST,][]{ivezic_lsst_2019}, which will be carried out over ten years with the Vera C. Rubin Observatory, will provide ground-based images of billions of galaxies. The cosmic shear results on this data are expected to significantly surpass the precision of existing surveys. In order to realize this potential, observational systematic effects for Rubin must be characterized and modeled at levels much stricter than stage-III experiments; for example, requirements on multiplicative shear bias for LSST are smaller by a factor of 4-5 \citep{LSST_SRD}. 

Observed images of objects, \eg galaxies, are the convolution of a true image by
the point-spread function (PSF) of the optical system, which describes the image of
a point source through the optics, sensors, and, in the case of ground-based
instruments, turbulence in the atmosphere. Therefore, accurately estimating galaxy
shears from observed images relies on a robust understanding of the PSF profile:
using a mis-estimated PSF during shape measurement leads to biased shear estimates.
In particular, errors in PSF shape (size) cause additive (multiplicative) shear bias
\citep{paulin-henriksson_point_2008, massey_origins_2013}. It is crucial to model not
only the PSF profile but also its spatial variation across the field of view, on each
exposure, to avoid introducing spatially correlated bias to the cosmic shear signal
\citep{rowe_improving_2010, heymans_impact_2012, jarvis_science_2016}. For a thorough
review on the topic of PSF modeling for weak lensing, see \cite{liaudat_point_2023}.

The three ground-based surveys mentioned above (KiDS, HSC, and DES) each used a
different PSF modeling method for their recent cosmic shear results: \texttt{lensfit}
\citep{miller_bayesian_2013}, \texttt{PSFEx} \citep{bertin_automated_2011}, and \piff
\citep{jarvis_dark_2020}, respectively. All performed within their survey requirements;
see \cite{jefferson_reanalysis_2025} for a systematic comparison between the KiDS-1000,
HSC-Y3, and DES-Y3 shear and PSF catalogs. More recently, modifications to \piff, including the addition of a color interpolation term, led to
improved performance on the DES-Y6 analysis compared to DES-Y3
\citep[hereafter \schutt]{schutt_dark_2025}. 
% , most notably the implementation of a spectral-dependent PSF correction
Shear measurements with LSST will also rely on PSF models from \piff, which is now
implemented within the \lsstpipe \citep{bosch_pipelines_2018, bosch_pipelines_2019, RubinPipe2025}.

Due to differences in everything from camera sensors and mirror configuration to local
topography, dome environment, and exposure time, the Rubin PSF will be different from the
DES PSF, and therefore \piff performance may also vary. More importantly, the tighter
bias requirements for LSST cosmic shear necessitate even more accurate PSF models than
were needed for DES. We aim here to provide an \textit{initial} evaluation of \piff
readiness, as configured in the \lsstpipe, for cosmic shear measurements with LSST.

In the absence of survey-quality LSSTCam data, as Rubin commissioning was still
underway at the time of this analysis, we instead estimate the accuracy of the PSF
modeling on simulated LSST-like images. With simulated images we are able to tailor
what effects we consider in our analysis. For instance, our atmospheric PSF simulation
uses local empirically-driven weather patterns which were shown to imprint anisotropic 
correlations in PSF shape and size across simulated LSST-sized exposures; the
anisotropy direction of these correlations persisted across many exposures due to the
prevailing wind over the observatory \citep{hebert_generation_2024}. In this work we
are able to test whether such atmospheric patterns are expected to hinder PSF modeling
for Rubin.

We run PSF modeling on these simulated LSST-like images and analyze the results
for evidence of modeling issues, such as the presence of static patterns in PSF parameter residuals. As there are no cosmic-shear-specific requirements on PSF modeling quality, and in
particular no set requirements on the spatial correlation of PSF residuals, we
propagate PSF residual statistics to an estimated contamination on the cosmic shear
data vector, in order to evaluate the PSF modeling quality quantitatively. We develop a framework to estimate the coefficients for this propagation, which link the PSF modeling errors to resulting shear bias, using simulations based on \cite{zhang_impact_2023}. We compare this
PSF contamination vector to a calculated estimate of cosmic shear uncertainty given
the simulated Rubin shear catalog from \cite{sheldon_metadetection_2023}.

We describe our simulated data products in \secref{sims} and \secref{mcal}, and
overview the PSF modeling with \piff in \secref{psf-modeling} and the cosmic shear
covariance estimate in \secref{cov}. We define our PSF diagnostics and the
propagation of PSF residuals to cosmic shear in \secref{measure}. We discuss our results and
their impact in \secref{results}, and outline takeaways and limitations in \secref{conclusions}.

\section{Simulations}\label{sec:sims}

We run two types of simulations in this work: slow, semi-realistic,
full-focal-plane images containing only stars, to perform the PSF modeling and
evaluation of residuals (see \secref{starsim}, \ref{sec:psfws}, and \ref{sec:opsim}); and
fast, approximate, postage-stamp coadd images of galaxies to connect our PSF
results to bias on cosmic shear (see \secref{galaxysim}). Specifically, we  use the galaxy simulations to obtain an expected PSF-to-galaxy size ratio distribution (\secref{coeff}) as well as inputs to the weak lensing covariance calculation (\secref{cov}).

\subsection{Star Simulations}\label{sec:starsim}

We use the \imsim\footnote{\url{https://github.com/LSSTDESC/imSim}} software
package to create semi-realistic images mimicking those produced by the Rubin
Telescope and LSST Camera.  The \imsim\ package uses the \galsim\ library to
generate various physical affects that determine the path of photons through
the system and to render images. Within \imsim\ are the definitions of the
components of the system, such as an approximate telescope optics (ray traced
using \batoid\footnote{\url{https://github.com/jmeyers314/batoid}}), filter
transmission curves, CCD detectors, atmosphere (see \secref{psfws}),
including differential chromatic refraction (DCR) effects.

In order to realize realistic physical effects, images are generated using ray
tracing through all components, including the atmosphere.  We do not simulate
bright stars ($i < 17$), which would require a large number of photons to
render and would dominate the computing time.

We developed the wrapper package
\montauk\footnote{\url{https://github.com/esheldon/montauk}} in order to
facilitate using the full set of features of \imsim\ as a python library rather
than using the \galsim\ configuration interface, and to have more control over
some features.  The \montauk library also introduces features that allow
additional experimental control, such as saving a two dimensional image
background and an accurate estimate of the effective world coordinate system
(WCS) mapping for the image. Saving these data allows us to avoid potential
errors that may occur when trying to determine them from the image, which could
then be conflated with algorithmic errors in the PSF determination code.

In order to assert additional control, we do not pass the images through the
simulated readout process, but use the electron count images directly. Nor do
we simulate detector defects or cosmic rays.  We did experiment with tree
rings (a type of detector imperfection that causes concentric patterns in pixel size; see \cite{esteves_photometry_2023}), but finding no significant effect on the PSF determination we do not
include them in the final set of simulations.
A discussion of several effects not included in these simulations, including detector realism, and their potential impact on results, can be found in \secref{conclusions}.

\subsection{Atmospheric PSF}\label{sec:psfws}

The atmospheric PSF simulation is implemented with \galsim\ as in
\cite{the_lsst_dark_energy_science_collaboration_lsst_2021}: prior to the
telescope aperture, photons are ray-traced through a set of 2D layers (or
``screens'') of spatially varying phase shifts which follow a \vonkarman\
turbulence spectrum \citep{von_karman_progress_1948}. These screens translate
horizontally across the aperture over the course of an exposure, approximating
the dynamic nature of the atmosphere.

Turbulence and wind profiles are used to weight the relative phase
contributions from each screen and to determine the speed and direction at
which the screens translate, respectively. We use \psfws\footnote{\url{https://github.com/LSSTDESC/psf-weather-station/}}
\citep{hebert_generation_2024}, a package that produces sets of wind and
turbulence profiles that are realistically correlated across altitudes. These
profiles are based on local measurements and weather forecast data products so
are representative of weather conditions at the chosen site. We use the
same inputs to \psfws\ as in \cite{hebert_generation_2024}: data from the
Gemini weather tower\footnote{
    $\sim$1km away from Rubin. At the start of the
    project, we did not yet have reliable Rubin weather tower data. That is now
    available. We note that while the dominant wind direction at Rubin differs
    from at Gemini due to local topography, the presence of a dominant wind matters
    more here than its specific direction. The spread of directions as well as the
    wind speed distributions are consistent between the two observatories.}
and ERA5\footnote{
    European Center for Medium-range Weather Forecasting ReAnalysis v5
    \url{https://www.ecmwf.int/en/forecasts/dataset/ecmwf-reanalysis-v5}}
forecasts\footnote{Closest grid point to Rubin (at -30.241, -70.737) is
 $\sim$1.9 km away at -30.25, -70.75.}.

The overall seeing and airmass for a given exposure are set by values from
the \opsim\ database, described in \secref{opsim}, and the \vonkarman turbulence  outer scale parameter $L_0$ for
each observation is drawn from a log normal with mean at 25\unit{m}, following \cite{the_lsst_dark_energy_science_collaboration_lsst_2021}.

\begin{figure}
    \includegraphics[width=0.47\textwidth]{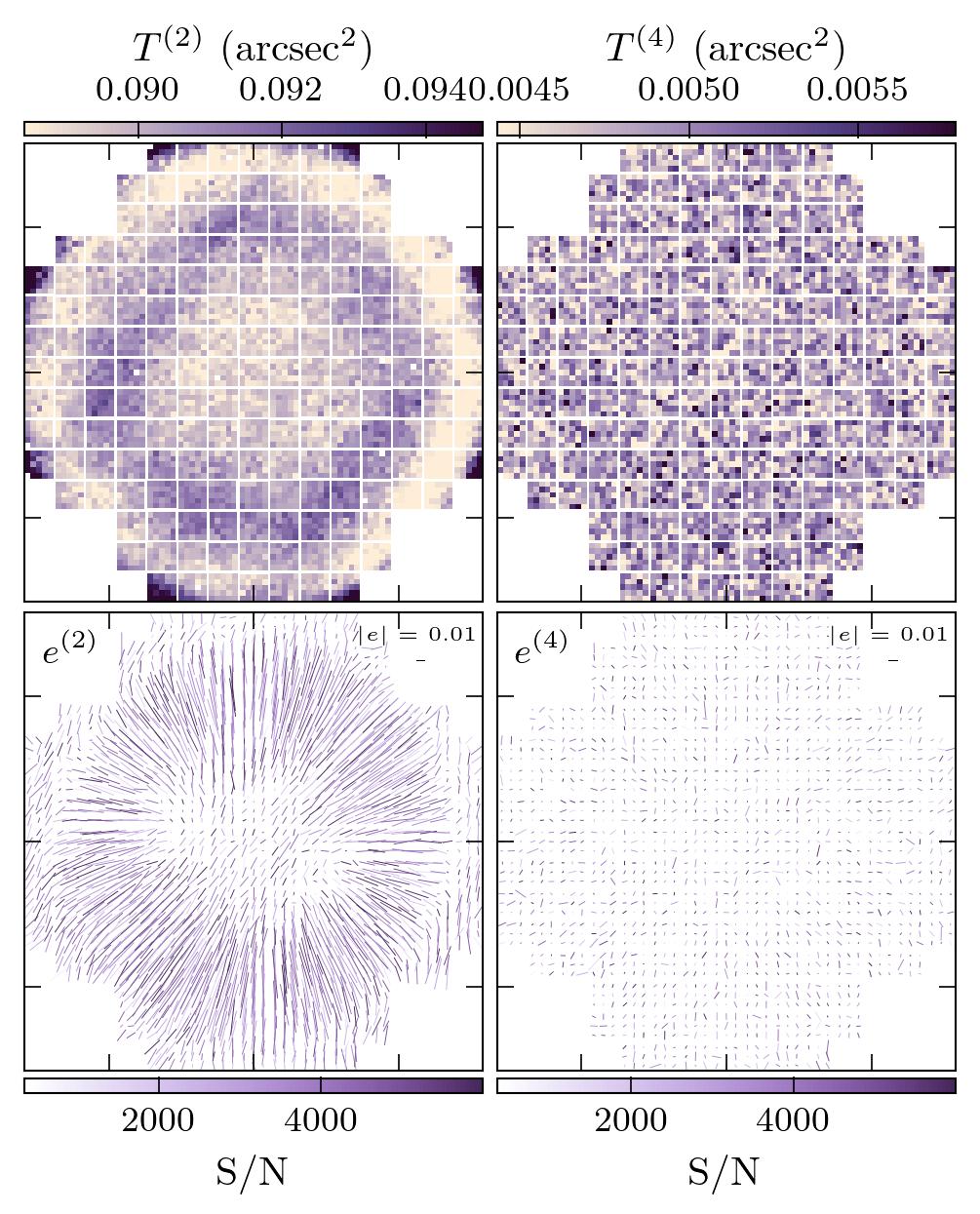}
    \caption{\label{fig:visit-example-low}
        PSF parameters for a low-seeing example
        exposure, FWHM\,$\sim$\,0.5\,arcsec; PSF size $\ttwo$ (upper
        left), shape $\gtwo$ (lower left) and fourth-order moment analogs $\tfour$
        (upper right) and $\efour$ (lower right). The length and orientation of
        the whiskers in the lower panels show the magnitude and orientation of
        the binned PSF ellipse, and whisker color encodes the combined star S/N 
        within each bin.\\
        }
\end{figure}

\begin{figure}
    \includegraphics[width=0.47\textwidth]{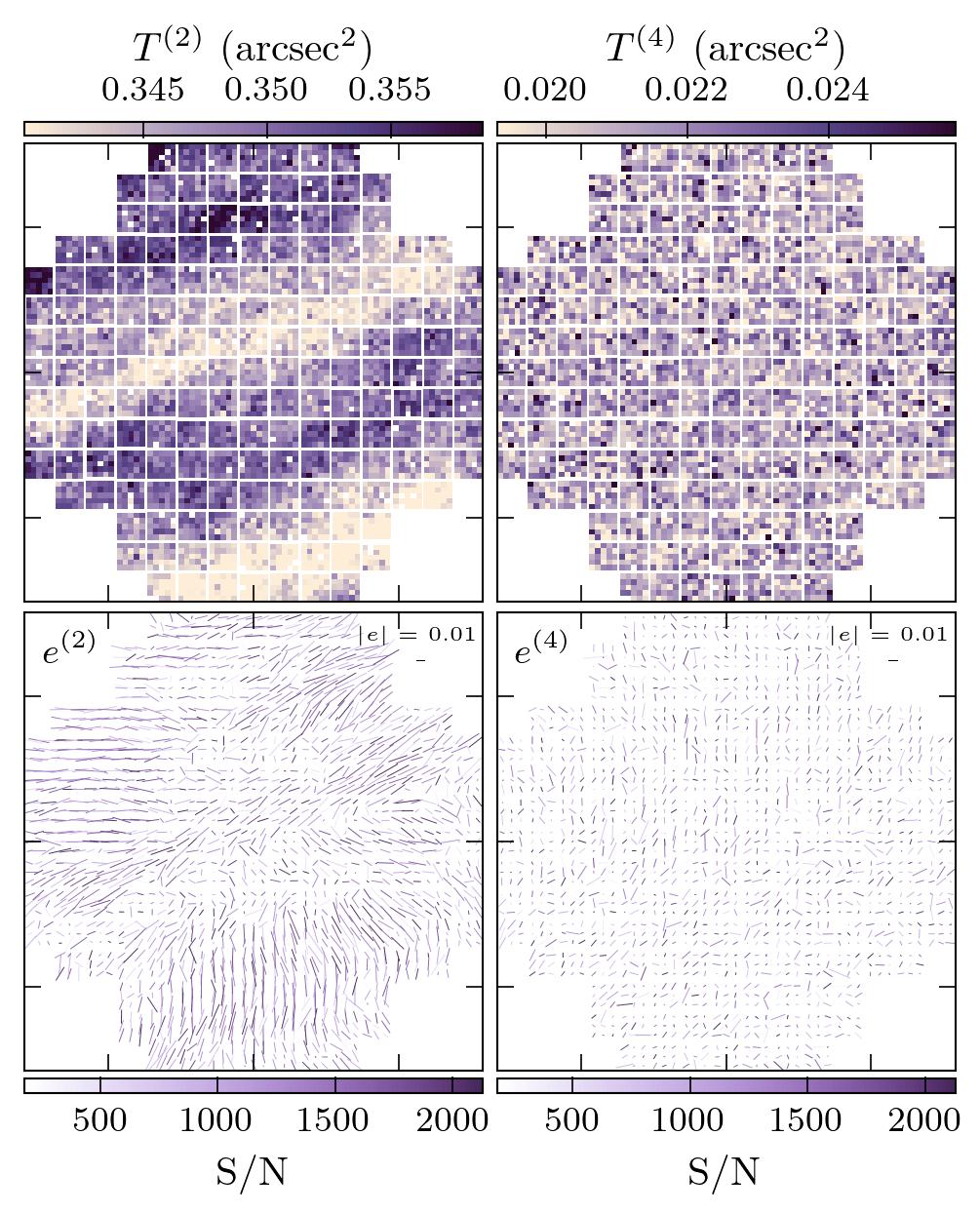}
    \caption{\label{fig:visit-example-high}
        Similar to \figref{visit-example-low} but for a high-seeing example exposure, FWHM\,$\sim$\,1\,arcsec.
        }
\end{figure}

We visualize PSF variation across the LSSTCam focal plane for example simulated
exposures in Figures~\ref{fig:visit-example-low} and~\ref{fig:visit-example-high}.
Each of the four panels shows measured values of a PSF parameter describing either
size ($T$) or shape ($e$), as defined in \secref{moments}. The annular optical
aberration implemented in \imsim\ is visible in the optics-dominated PSF of the
low-seeing visit in \figref{visit-example-low}; in contrast, the PSF of the
high-seeing visit in \figref{visit-example-high} is dominated by characteristic
atmospheric patterns caused by turbulence blowing across the aperture.

Across our 1508 simulated exposures we observe an anisotropy in PSF ellipticity:
the distribution of PSF orientation angle peaks at the same on-sky orientation as
the distribution of input ground wind directions. This is consistent, as expected
given the common use of \psfws, with the correlation found between wind and PSF
anisotropic two-point statistics in \cite{hebert_generation_2024}. We also find 
near-coincident distributions of PSF orientation angle and wind direction in a
subset of visits of the DES-Y3 PSF catalog for which we have associated wind
measurements, potentially implying that a persistent wind-driven anisotropy is
present in the DES PSF, \ie, may be realistic to expect for Rubin.

\subsection{Operation Simulation and Star Catalogs}\label{sec:opsim}

The simulated observations are drawn from a database generated by the
Operation Simulation (\opsim) code\footnote{\url{https://rubin-sim.lsst.io/}} \citep{connolly_end_2014, yoachim_optical_2016} for the full 10 years of the survey.
We use version ``baseline 3.3'' of the database for this work.  The database
provides sky pointings as well as the relevant observing conditions such as
telescope altitude and azimuth, location of the moon, sky background and
seeing. LSST cosmic shear will be measured on the wide-fast-deep
(WFD) survey; for our simulations we use 1508 $i$-band pointings covering
a $\sim$100 square degree patch of the WFD area. The input star catalogs
are the same as used for the Data Challenge 2 (DC2) simulation project
\citep{the_lsst_dark_energy_science_collaboration_lsst_2021}.  As mentioned
above, we do not draw stars with $i < 17$, but otherwise the rendering was the
same as used for DC2.

\subsection{Galaxy Simulations}\label{sec:galaxysim}

PSF determination errors will propagate into weak lensing shear measurements.
In order to predict this contamination, we need simulations that include
galaxies.  However, rendering the high density of galaxies expected in LSST
data with \imsim\ would exceed our computing resources.  We instead use a
simpler simulation with analytic PSFs tuned to approximate that provided by
\imsim\ and \psfws.  The density of pointings is matched to the \opsim
database described in \secref{opsim}.

The PSF for these images is modeled as three separate components: an
atmosphere, the optics, and local diffusion effects on each CCD.

For the atmospheric part of the PSF, we use the ``power spectrum psf'' from
the \descwlshearsims package \citep{mdetlsst2023}.  This PSF was tuned
to have mean FWHM of 0.625 arcsec, with isotropic spatial variation in FWHM
and ellipticity consistent with measurements from \cite{heymans_impact_2012}.

The optical part is inspired by the realistic optical PSF produced by \imsim\
package\footnote{\url{https://github.com/LSSTDESC/imSim}}.  The spatial
variations are modeled using order 6 Zernike polynomials \citep{Zernike1934},
constrained to have circular symmetry.  The profile is modeled as an
elliptical Moffat \citep{Moffat1969}.  The optical model is tuned to give a
mean FWHM of about 0.35 arcsec.

Finally, each CCD is given a random, fixed Gaussian diffusion kernel, with
FWHM drawn from a normal distribution with mean of 0.2 arcsec and
$\sigma=0.02$.

These three PSF components are tuned to provide a mean seeing FWHM of
approximately 0.8 arcsec and variations $\sigma=0.06$ arcsec, similar to
the $i$-band \imsim images.  Note, due to the non Gaussian shape of the optical and atmospheric
profiles, the individual FWHM quoted above do not add quadratically to give 0.8.

To generate galaxies, we use the \descwlpackage package
\citep{DESCWLSanchez2021}\footnote{\url{https://github.com/LSSTDESC/WeakLensingDeblending}}.
Each galaxy has three components:  a bulge component, represented by a de
Vaucouleurs' profile \citep{devauc1948}, a disk represented by an exponential
profile, and an active galactic nucleus represented as a point source.

For efficiency we do not render full images, but rather individual postage
stamps for each object and PSF that are then coadded.  The PSFs are rendered in the
coadd pixel coordinates to avoid image warping, which would dominate the
processing time.

We anticipate that LSST shear will be measured on $riz$ coadded images \citep[as in DES-Y6][]{yamamoto_dark_2025};
therefore, each coadd contains the expected total number of epochs $N_{\rm
epoch}^{riz}=460$ for year 10 (Y10). We determine this number by combining the
$i$-band epochs found in our simulated field, according to the \opsim\
pointings, with the number of epochs in $r$- and $z$-band, relative to
$i$-band, from the LSST Science Requirements
Document\footnote{\url{https://docushare.lsst.org/docushare/dsweb/Get/LPM-17}}.  These simulations are repeated with $N_{\rm
epoch}^{riz}=46$ for year 1 (Y1).

\section{Metacalibration Measurements of Galaxy Simulations}\label{sec:mcal}

We process the galaxy simulations described in \secref{galaxysim} using
\metacal\footnote{It is worth clarifying that LSST processing will run \metadet for galaxy shear measurement, rather than \metacal. For the isolated objects processed in this analysis, the difference is trivial. \label{mcalvmdet}}\citep{SheldonMcal2017}.  During the \metacal processing, the image is
deconvolved by the PSF and reconvolved by a round Gaussian kernel slightly
larger than the original PSF. For the shape measurements we fit a gaussian to
the PSF and galaxy, following the same procedure as the Dark Energy Survey year
3 processing \citep{gatti_dark_2021}.  This results in a PSF size measured for
each object $T$, which is the trace of the Gaussian covariance matrix, as well
as pre-PSF $T_{\rm gal}$ and reduced shear shape parameters $g_1, g_2$
for each galaxy. In addition to these parameters we measure weighted moments
as described in \secref{moments}.

\section{PSF modeling}\label{sec:psf-modeling}
In overview, PSF modeling relies on measurements of the PSF at locations of
stars in a particular image. A PSF solution at the location of other objects
is found by interpolating between these PSF observations. Typically, some
``reserve'' stars are withheld from those ``PSF'' stars used to
build the interpolation, and serve as test points with which to evaluate
its performance. PSF modeling choices include: star selection criteria,
a parameterized model to describe the PSF profiles, a functional form for the
interpolation of those PSF parameters, \etc\

\piff\footnote{\url{https://github.com/rmjarvis/Piff}} (PSFs in the Full
Field-of-View) \citep{jarvis_dark_2020} is a state-of-the-art PSF modeling
package for ground-based imaging data. Here we give an overview of \piff\ as
configured in the \lsstpipe\footnote{\url{https://github.com/lsst/meas_extensions_piff}} \citep[see Sec. 5.7 in ][]{RubinPipe2025}, which
we use in this work, and refer the interested reader to \schutt\ for a
thorough explanation of \piff\ and all its capabilities. For full parameter
defaults, see the \texttt{piffPsfDeterminerConfig} documentation for
\lsstpipe\ release \texttt{w\_2024\_46}. We run PSF modeling on our
single exposure star-only simulations (described in Sections
\ref{sec:starsim}, \ref{sec:psfws}, and \ref{sec:opsim}).

We perform the default star/galaxy separation on our catalogs to select stars
for PSF modeling; this removes any blended stars and ensures we use the same
object selection as in real data. After removing parents and blends, the default
star/galaxy separation includes a selection on objects with \texttt{base\_PsfFlux\_instFlux}\,$>$\,12500, S/N\,$>$\,50.
After this selection and \piff\ outlier rejection, on average per detector
we have $\sim$110 PSF stars for fitting and an additional $\sim$30 reserve stars
for testing. No detector images with fewer than 48 stars are used.

The profile of each individual star is modeled with a \texttt{PixelGrid}:
a 2D grid of what \schutt\ calls ``model pixels'' that are a continuous,
kernel-smoothed representation of the postage stamp pixel values. The
value of each model pixel is interpolated between stars across the image with
a second-order polynomial. This interpolation is performed per CCD image, in
CCD pixel coordinates. At the time of our tests the implementation of this
interpolation in sky coordinates was not ready for use in the \lsstpipe.

% model parameters:
% modelSize=25,
% interpolant=lanczos(11)

% spatial interp parameters:
% stampSize=35,
% order=2,
% type='BasisPolynomial',
% coord=pixel

% outliers:
% type = Chisq
% nsigma = 4

\begin{figure}
    \includegraphics[width=0.47\textwidth]{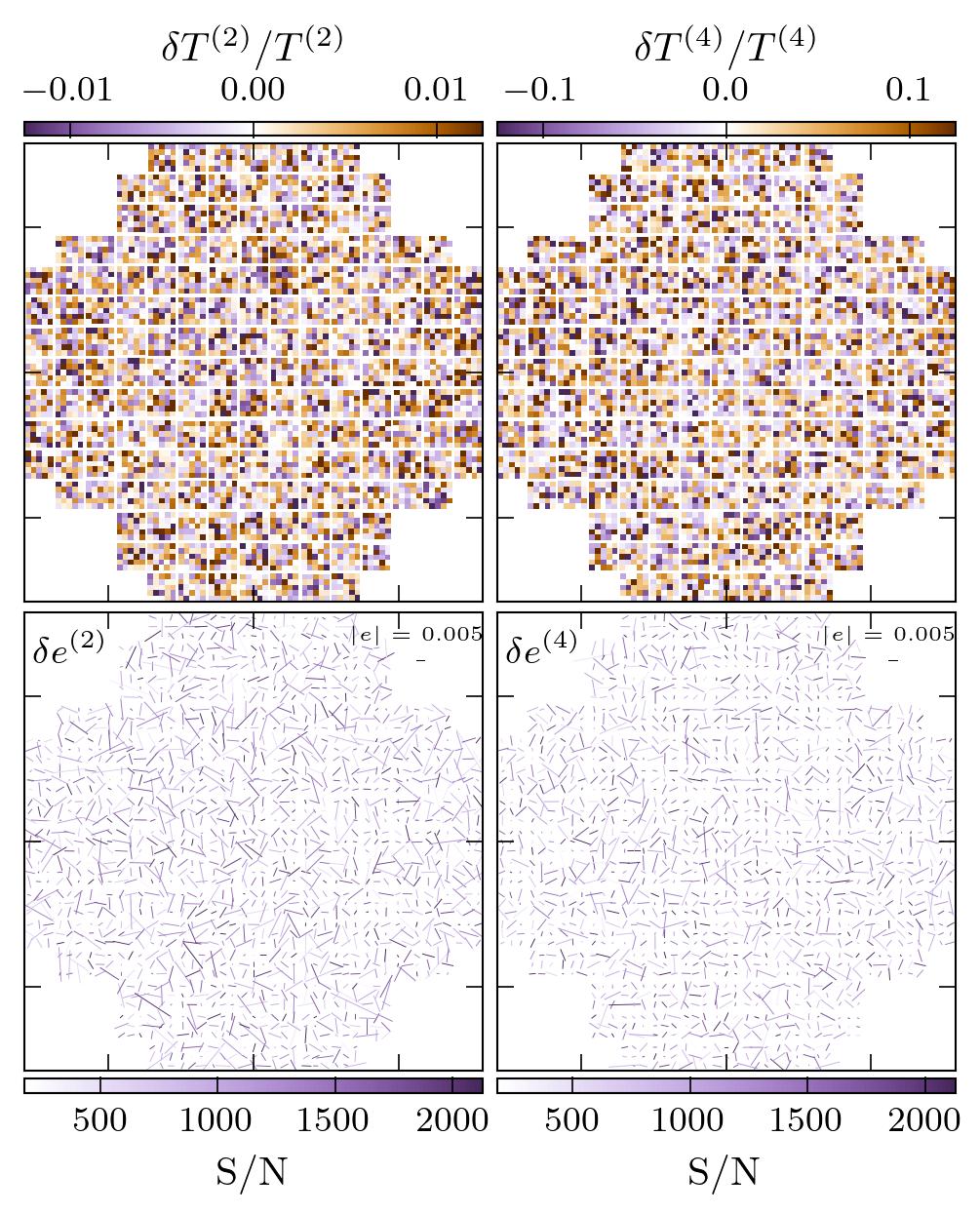}
    \caption{\label{fig:visit-example-resid}
        PSF parameter residuals for the high-seeing example exposure shown in
        \figref{visit-example-high}; PSF size $\delta \ttwo/\ttwo$ (upper left),
        shape $\delta \gtwo$ (lower left) and fourth-moment analogs $\delta\tfour/\tfour$
        (upper right) and $\delta\efour$ (lower right). The length and orientation of the
        whiskers show the magnitude and orientation of the residual PSF ellipse.
        }
\end{figure}

The PSF size, shape, and weighted moments (see \secref{measure}) are measured
with \ngmix\footnote{\url{https://github.com/esheldon/ngmix}} on postage
stamp images drawn from the \piff\ models, for both the observed stars and the
PSF model prediction.
The residuals of the PSF parameters (defined in \secref{moments}) are a measure
of \piff\ model accuracy. The residuals for the high-seeing example exposure (see \figref{visit-example-high}) are shown in \figref{visit-example-resid}. No evidence of the focal-plane-wide atmospheric structure remains; as well as not
showing signs of systematic orientation, the whiskers are roughly half the size
as in \figref{visit-example-high}.

\section{Theoretical weak lensing covariance}\label{sec:cov}
In order to assess the statistical importance of the residual
biases caused by PSF modeling, we compute a realistic cosmic shear covariance matrix that
accurately describes the statistical precision we can expect from the upcoming LSST
data releases. The matrix is computed analytically, following the methods described
in \citet{cosmoSLICS} and used in e.g. \citet{KiDS1000_Joachimi, reischke_kids_2025}\footnote{\url{https://onecovariance.readthedocs.io/}}. We repeat the
calculation for the two scenarios presented in this paper, namely the LSST
Y1 case, with (unweighted) $n_{\rm eff}=17.5$ and $\sigma_{\epsilon}=0.29$, and
the Y10 case, with $n_{\rm eff}=34.3$, $\sigma_{\epsilon}=0.27$. These parameters,
along with redshift distributions, are calculated from the simulated galaxy
catalogs described in \secref{galaxysim} and \ref{sec:mcal}. The calculations
assume a survey footprint 19,600 deg$^2$. The cosmology matches that of the
{\it OuterRim} Simulation \citep{OuterRim}, a Tera-particle $N$-body simulation
that serves as the backbone of large-scale structures for the flagship cosmo-DC2
catalogs~\citep{cosmoDC2}. Namely, it assumes a flat $\Lambda$CDM cosmology
universe with a matter density of $\Omega_{\rm m} = 0.2648$, a Hubble parameter of
$h=0.71$, a baryon density of $\Omega_{\rm b} = 0.0448$, a tilt in the primordial
power spectrum of $n_{\rm s}=0.963$, and a normalization of the matter fluctuations
set to $\sigma_8=0.801$.

The covariance matrix contains contributions from the Gaussian term, the connected
non-Gaussian term and the super-sample covariance terms, and therefore offers an
accurate estimate of the sample variance and of correlations between the different
parts of our data vectors. It assumes a power spectrum calculated from
\textsc{halofit} \citep{Halofit2012}, no intrinsic alignments, and a perfectly
circular footprint. These are approximations that affect the size of our error
bars by a few percent at most, hence are fine to adopt. We refer the reader to
\citet{cosmoSLICS} for full details on the calculations.

\section{PSF metrics}\label{sec:measure}
There are many well-documented PSF diagnostics tests in the weak lensing
literature; here we present a subset of those metrics that are relevant for
our simulation case. We define these based on a series of recent studies,
namely~\cite{zhang_impact_2021,zhang_impact_2023}, and the PSF modeling
and shear systematics analysis papers of Hyper Suprime-Cam Year 3
\citep[hereafter \zhang]{zhang_general_2023} and the Dark Energy Survey Year
6 \citep[\schutt, ][]{yamamoto_dark_2025}. These works defined fourth-order PSF
moments, and, by including them in weak lensing-related metrics, found their
impact on galaxy shear measurements to be as, or more, significant than the
traditional second moments.
%this last is basically just paraphrased from \schutt...

Our definitions of PSF size and shape parameters, and analogous fourth-order
quantities, are presented in \secref{moments}; two-point statistics of these PSF
parameters and their model residuals, and the total additive PSF
contribution to cosmic shear, are defined in \secref{2pt-formalism}. In
\secref{coeff} we approximate the additive shear bias contributions from
specific types of PSF errors.

\subsection{Moments}\label{sec:moments}

First, we define normalized weighted moments $M^{\omega}_{pq}$\footnote{Moment order $n=p+q$, as in \zhang.},
\begin{eqnarray}\label{eqn:moments}
    M^{\omega}_{pq} = \frac{1}{M^{\omega}_{00}}\int dx \, dy \, x^p \, y^q \, \omega(x,y) \, I(x,y) \,,
\end{eqnarray}
for a centered image $I(x,y)$, weight $\omega(x,y)$, and with flux normalization
$M^{\omega}_{00} = \int dx \, dy \,\omega(x,y) \, I(x,y)$. From these moments, we construct
parameters to describe the size and shape of an intensity profile, \eg, the PSF.
% or introduce spin-0 spin-2 here to generalize to 4th moms?

Starting with second-order moment quantities, we use the moment trace $T$ and distortion $e$, a spin-2 complex quantity $e=e_1+ie_2$. We define these parameters in terms of \textit{deweighted}\footnote{
    Moment deweighting is standard for galaxy shear measurements, if not often made explicit. Weighted moments are biased estimates of galaxy shear, so one must remove the effect of the weight.
    This is implemented in \ngmix as $\mathbf{M} = \left((\mathbf{M^{\omega}})^{-1} - \mathbf{W}^{-1}\right)^{-1}$, where $\mathbf{M}$, $\mathbf{M^{\omega}}$, and $\mathbf{W}$ are the deweighted, weighted, and weight second moment matrices. 
    }
adaptive moments $M_{pq}$ \citep{bernstein_shapes_2002}:
\begin{align}
    \ttwo &= M_{20}+M_{02} \,,\\
    \gtwo &= \frac{M_{20}-M_{02} + i\,2M_{11}}{M_{20}+M_{02}} \,,
\end{align}
where the superscript $(2)$ specifies that these are parameters constructed from
second moments. 

The size and shape parameters $\ttwo$ and $\gtwo$ are, respectively, spin-0 and
spin-2 combinations of second moments; \zhang\ and \schutt\ constructed
analogous spin-0 and spin-2 parameters from fourth-order moments. We define
$\tfour$ and $\efour$ as follows, equivalent to \schutt\footnote{
    Compared to \zhang's fourth-order shape $M^{(4)}_{\rm PSF}$ and size
    (kurtosis) $\rho^{(4)}_{\rm PSF}$, we find $\efour=M^{(4)}_{\rm PSF}$ and
    $\tfour = (M^{\omega}_{20}+M^{\omega}_{02})\rho^{(4)}_{\rm PSF}$.
}:
\begin{align}
\tfour &= \frac{M^{\omega}_{40}+2M^{\omega}_{22}+M^{\omega}_{04}}{M^{\omega}_{20}+M^{\omega}_{02}}  - \ttwo \,, \label{eqn:t4} \\
    \efour &= \frac{M^{\omega}_{40}-M^{\omega}_{04} + i\,2(M^{\omega}_{31}+M^{\omega}_{13})}{(M^{\omega}_{20}+M^{\omega}_{02})^2}- 3 e^{(2)}\,. \label{eqn:e4}
\end{align}
These are equivalent to the $\tfour$ and $\efour$ parameters defined in \schutt.
The subtraction of $\ttwo$ and $e^{(2)}$ in \eqnref{t4}, \ref{eqn:e4} serves to
decorrelate the fourth-order parameters from those of second-order; \schutt\ notes
that $\tfour$ and $\efour$, as defined above, are nearly zero for an elliptical Gaussian.

To clarify our notation of spin-2 quantities, we use $e$ for distortion and $g$
for reduced shear/ellipticity definitions of shape \citep[see Eq 2-7, 2-8
respectively in][]{bernstein_shapes_2002}. Our $\gtwo$ is identical to the $e$
\textit{distortion} used in \zhang, and is different from the $e^{(2)}$
\textit{ellipticity} used in \schutt. However, this notation does not distinguish
that $\gtwo$ is an adaptive (\ie, deweighted) moment parameter whereas $\efour$
is defined in terms of weighted moments.

We define the model residual for parameter X as $\delta X = X_* - X_{\rm model}$,
where the ``$*$'' subscript indicates the PSF parameter measured on data, and
``model'' on the \piff\ model prediction. To simplify notation we write
$X_{\rm model}$ as $X$, unless otherwise stated. For example, we write the
fractional residual $\frac{\delta X}{X_{\rm model}}$ as $\frac{\delta X}{X}$.

\subsection{Two-point statistics of PSF moments}\label{sec:2pt-formalism}
The PSF can contaminate galaxy shear measurement with an additive error term,
\ie, $g_{\rm obs}\simeq g_{\rm gal} + \delta g^{\rm PSF}_{\rm sys}$. This
contamination can come from both PSF leakage (if the shear measurement algorithm
introduces some dependence on the PSF) and via modeling errors (if the PSF given
to the shear measurement algorithm is incorrect). This has previously been
expressed as a linear combination of PSF second moment parameters and residuals
in~\cite{paulin-henriksson_point_2008,rowe_improving_2010,jarvis_science_2016},
and more recently, \zhang\ introduced contributions from fourth-order PSF moments.
We will follow~\cite{yamamoto_dark_2025}:
\begin{align*} \label{eqn:delta-g}
    \delta g^{\rm PSF}_{\rm sys} &= \alpha_2 \gtwo + \beta_2 \delta \gtwo + \eta_{22} \gtwo \frac{\delta \ttwo}{\ttwo} \\
    &+ \alpha_4 \efour + \beta_4 \delta \efour  + \eta_{44} \efour \frac{\delta \tfour}{\tfour} \numberthis \\
    &+ \eta_{24} \gtwo \frac{\delta \tfour}{\tfour} + \eta_{42} \efour \frac{\delta \ttwo}{\ttwo}\,.
\end{align*}
These terms represent the total PSF contamination to the measured galaxy shear,
so each term is a spin-2 quantity; the PSF size error $\frac{\delta T}{T}$ is a
scalar, but it contributes a shape error proportional to the shape of the PSF,
\eg, a size error on a round PSF contributes zero additive shape error.
The $\alpha, \beta, \eta$ coefficients determine the extent to which any PSF
parameter or residual propagates to the measured galaxy shear; see
\secref{coeff} for more details.

We can write \eqnref{delta-g} more concisely as
\begin{equation}
    \delta g^{\rm PSF}_{\rm sys} = \sum_{k=1}^{8} c_k P_k \,,
\end{equation}
where $\mathbf{c}=[\alpha_2, \beta_2, \eta_{22}, \alpha_4, \beta_4, \eta_{44}, \eta_{24}, \eta_{42}]$
is the vector of PSF contamination coefficients, and
$\mathbf{P}=[\gtwo_{\rm model}, \delta\gtwo, w_{22}, \efour_{\rm model}, \delta\efour, w_{44}, w_{24}, w_{42}]$
is the vector of spin-2 PSF parameters and residuals. We have further simplified our
notation by defining
\begin{equation}\label{eqn:wij}
    w_{ij}\equiv e^{(i)}\frac{\delta T^{(j)}}{T^{(j)}}
\end{equation}

The additive bias $\delta g^{\rm PSF}_{\rm sys}$ on an individual shear measurement
can propagate to additive bias on a measurement of cosmic shear, the two-point
correlation function of galaxy shears $\xi_{\pm} = \langle g_{\rm obs}g_{\rm obs} \rangle$.
The additive bias on $\xi_{\pm}$ from PSF contamination is then
\begin{align*}
    \delta \xi_{\pm}^{\rm PSF} &= \langle \delta g^{\rm PSF}_{\rm sys}\delta g^{\rm PSF}_{\rm sys} \rangle \\
    \numberthis &=\sum_{k=1}^{8}\sum_{l=1}^{8} c_k c_l \langle P_k P_l \rangle \,. \label{eqn:dxi-psf}
\end{align*}
The $\langle P_k P_l \rangle$ are two-point functions of PSF parameters; these
36 correlation functions (symmetric in $k,l$) are collectively referred to as rho statistics, and
describe the correlation, as a function of separation angle $\theta$, of pairs
of PSF parameters and/or residuals. These PSF-PSF correlation functions are a
common diagnostic criteria with which to evaluate PSF model performance, but,
without the $c_k c_l$ coefficients, cannot be interpreted as a contamination
to cosmic shear.

\subsection{Coefficients of PSF cosmic shear contamination}\label{sec:coeff}
The PSF contamination coefficients ($\alpha$, $\beta$, $\eta$ in \eqnref{delta-g})
are needed to compute an estimate of $\delta \xi_+^{\rm PSF}$. We find intuition
about the behavior of these coefficients in \cite{paulin-henriksson_point_2008}, who
predict that $\beta_2$ depends on the PSF-to-galaxy-size ratio as $\beta_2 \simeq \frac{1}{2} \left< T_{\rm PSF} /T_{\rm gal} \right>$\footnote{For simplicity we write $T$ for $\ttwo$ wherever this ratio appears.} for an unweighted moment measurement of shear $g$ for a Gaussian galaxy and PSF; under
these assumptions, \cite{jarvis_science_2016} find the same dependence for $\eta_2$.
The PSF leakage coefficients $\alpha$, by contrast, are set primarily by the shear measurement
algorithm's treatment of the PSF and its deconvolution. Current state-of-the-art shear measurement algorithms, such as \metadet
\citep{sheldon_mitigating_2020} and \texttt{AnaCal}
\citep{li_differentiable_2023, li_analytical_2025}, are expected to give $\alpha=0$;
indeed,~\cite{yamamoto_dark_2025} find $\alpha_2, \alpha_4$ consistent with zero
in the DES-Y6 \metadet shear catalog (see \tabref{coefficients}). 

In reality, the $\beta, \eta$ coefficients may also depend on choice of shear measurement algorithm, as well as on galaxy and PSF morphology. The PSF contamination coefficients in modern cosmic shear analyses are therefore fit from the data
by cross-correlating associated shear and PSF catalogs~\citep{hamana_cosmological_2020, gatti_dark_2021}.
This fit assumes that the coefficients are constant across the ensemble of galaxies, \eg,
$\langle \alpha_2 \gtwo \beta_2 \delta \gtwo \rangle = \alpha_2\beta_2 \langle
\gtwo\delta \gtwo \rangle$, although sometimes the fit is done separately for
each galaxy redshift bin.
In this analysis we do not estimate coefficients from our semi-realistic
simulations; the computational costs of including galaxies in our simulated
single visits, and co-adding the visits in order to measure galaxy shapes, would
be out of scope for this project. 
Neither do we directly use the \cite{paulin-henriksson_point_2008, jarvis_science_2016} model, which depends on unmet assumptions and cannot be extended analytically to higher-order moments.

Instead, we estimate the $\beta$ and $\eta$ parameters in simulation via the response of measured galaxy shape to injected PSF errors (more detail below), and in our fiducial analysis assume the a priori null values for \metadet $\alpha_2, \alpha_4$ (the impact of this choice is explored in the Appendix).

The $\beta$ and $\eta$ simulations are based on the study of shear response to
higher-order moments presented in \cite{zhang_impact_2023}: we pass a noiseless image of a Gaussian galaxy convolved by a Gaussian PSF to \texttt{metacalibration}\footref{mcalvmdet}, along with an error-added PSF to use during measurement.
The PSF error is injected into the image by iteratively solving the shapelet decomposition $\Delta b_{jk}$, where $j,k$ are the shapelet coefficients \citep[see Eq.~13 of ][]{zhang_impact_2023}, to create the desired PSF moments residual $\Delta M_{pq}$
\begin{equation}
    \sum_{j,k} \frac{\partial M_{pq}}{\partial b_{jk}} \Delta b_{jk} = \Delta M_{pq},
\end{equation}
where $\frac{\partial M_{pq}}{\partial b_{jk}}$ is the Jacobian matrix between the moments and the shapelet coefficients. In this work, we use shapelet decomposition up to the 4th order, i.e., $j+k\leq 4$.
For radial moments defined as a combination of higher moments, e.g., $T^{\rm (4)}(M_{40},M_{04},M_{20},M_{02})$ in \eqnref{t4}, we modify the PSF moments by a multiplicative factor $\delta m$
\begin{equation}
    \Delta M_{pq} = \delta m M_{pq}
\end{equation}
to ensure that $\Delta M_{pq}$ do not have non-spin-$0$ components.
For error added to each PSF parameter $P_k$ we measure the response of galaxy shape
error $\delta g$, the difference between true and \metacal-estimated galaxy shape, as a
function of $T_{\rm PSF}/T_{\rm gal}$.

\begin{figure}
\includegraphics[width=0.47\textwidth]{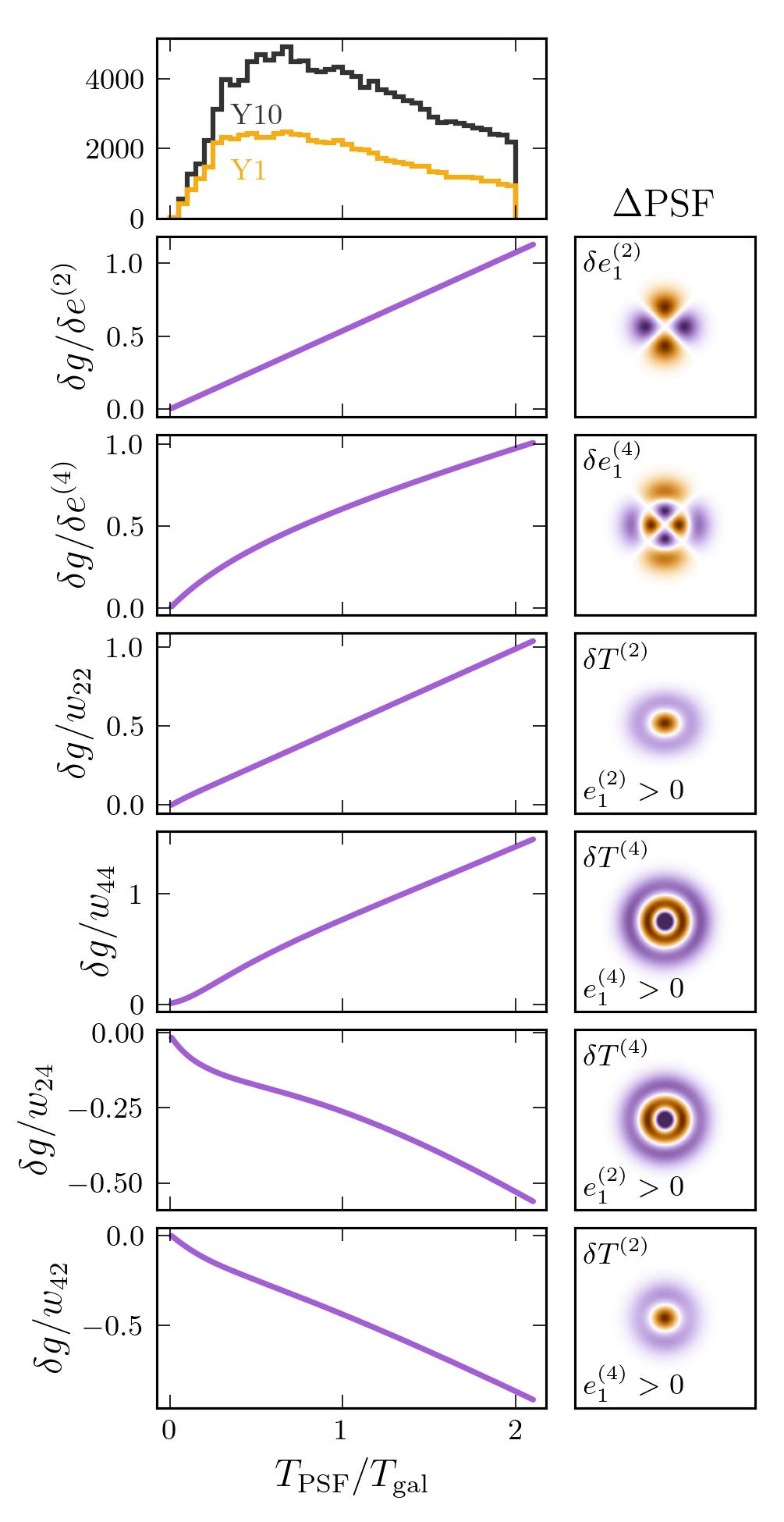}
\caption{\label{fig:mom-response}
    The response of galaxy shear bias to injected error in specific PSF
    parameters, as a function of the PSF-to-galaxy size ratio (left panels), for
    Gaussian galaxy and Gaussian PSF. Right panels show (exaggerated) change in the
    PSF profile with the addition of error in the labeled PSF parameter.
    Top panel shows the distributions of the PSF-to-galaxy size ratio for the
    Y1 (gold) and Y10 (dark grey) galaxy simulations described in \secref{galaxysim}.
    }
\end{figure}

The resulting galaxy response curves are shown in the left panels of \figref{mom-response}. The right column
shows an example change in the PSF profile for the type of PSF error added in each row;
size errors (bottom four rows) are added to an elliptical PSF in order to produce a
desired spin-2 response. 
The shear response curves for $\delta \gtwo$ and $\delta w_{22}$ are linear with a slope of $\sim$\,0.5, matching closely the predictions from \cite{paulin-henriksson_point_2008, jarvis_science_2016}. The shear responses to error in higher-order moment PSF parameters also follow a similar, though only approximately linear, trend with $T_{\rm PSF}/T_{\rm gal}$; $\delta w_{24}, \delta w_{42}$, however, have negative slopes.

To determine PSF contamination coefficient $c_k$ we integrate
$\delta g / \delta P_k$ over the $T_{\rm PSF}/T_{\rm gal}$ distributions in our
simulated galaxy catalogs (top panel of \figref{mom-response}, see \secref{galaxysim});
the coefficients values, estimated for both Y1 and Y10, are listed in \tabref{coefficients}.
This method of estimating $\beta$ and $\eta$ allows us to account for the dependence
of these coefficients on the relative PSF and galaxy sizes in a shear catalog. Y10 is
deeper than Y1, so we expect smaller galaxies relative to the PSF, which on average are
\textit{more} sensitive to PSF errors---this is reflected in the larger (magnitude)
PSF contamination coefficients found for Y10 relative to Y1. 
Due to this dependence on the shear catalog, in addition to potential differences
in PSF treatment between shear measurement algorithms, PSF contamination coefficient
values are in general not directly comparable between surveys.

\begin{table}
    \centering
    \begin{tabular}{cccc>{\color{gray}}c>{\color{gray}}c}
        \toprule
        & Y1 & Y10 & DES-Y6 & DES-Y6 fit\footnote{$\beta_2, \eta_{22}, \eta_{24}$ are halved to convert to distortion definition of $e^{(2)}$.} & HSC-Y3 fit\footnote{Values adjusted to LSST-like $\left< T_{\rm PSF}/T_{\rm gal} \right>$ (see text), and multiplied by $-1$ to match residual sign convention.}\\
        \midrule
        $\beta_2$   & 0.50    & 0.53    & 0.54    & 0.37 $\pm$ 0.29    & 0.56 $\pm$ 0.14       \\
        $\eta_{22}$ & 0.46    & 0.49    & 0.50    & 1.22 $\pm$ 3.6     & -                   \\
        $\beta_4$   & 0.55    & 0.58    & 0.59    & 1.95 $\pm$ 0.27    & $-$0.56 $\pm$ 0.14   \\
        $\eta_{44}$ & 0.70    & 0.73    & 0.75    & 3.15 $\pm$ 2.6     & -                   \\
        $\eta_{24}$ & $-$0.26 & $-$0.27 & $-$0.28 & $-$0.64 $\pm$ 0.34 & -                   \\
        $\eta_{42}$ & $-$0.41 & $-$0.44 & $-$0.45 & $-$10.8 $\pm$ 18.9 & -                   \\
        \addlinespace[.4em]
        $\alpha_2$  & 0       & 0       & -       & 0.005 $\pm$ 0.005    & - \\ %$-$0.02 $\pm$ 0.003
        $\alpha_4$  & 0       & 0       & -       & $-$0.006 $\pm$ 0.015 & - \\ %$-$0.17 $\pm$ 0.01
        \bottomrule
    \end{tabular}
    \caption{\label{tab:coefficients}
        Values of coefficients in \eqnref{delta-g}, determined from simulated
        shear moment responses (black; see \secref{coeff}) or from fits to DES and HSC
        datasets (grey; quoted respectively from~\cite{yamamoto_dark_2025} and \zhang). Y1 and Y10 values of $\alpha_i=0$ are assumed.
    }
\end{table}

It is reasonable, though, to predict DES-Y6 contamination coefficients with our method from the DES-Y6 size ratio distribution, given the use of \metadet for the DES-Y6 shear catalog and \metacal in our shear response curves. These predicted coefficients are listed in the black-text ``DES-Y6'' column of \tabref{coefficients}, and we compare them to those presented in~\cite{yamamoto_dark_2025}, reproduced in the grey-text ``DES-Y6 fit'' column.
The predicted $\beta_2=0.54$ is consistent with the
measured $\beta_2=0.37\pm0.29$, but the predicted $\beta_4$ is significantly smaller than the measurement.
Predictions of $\eta$ are consistent to roughly $1\sigma$ with measured values, though the measurements of $\eta_{22}, \eta_{42}$ are poorly constrained. The significant measurement of $\eta_{24}<0$ lends confidence to the negative shear response slope in \figref{mom-response}.
The predicted DES-Y6 coefficients are within $2.5\%$ of the LSST-Y10 values, consistent with the $\sim2\%$ difference in
$\left<T_{\rm PSF}/T_{\rm gal} \right>$ between the two catalogs.

Comparison with HSC-Y3 coefficients (\zhang) is less straightforward: HSC-Y3
shears are estimated with \texttt{reGauss}, which treats the PSF differently than
\metacal. In particular, the HSC $\alpha$ values are not relevant to our comparison
because we expect nonzero PSF leakage with \texttt{reGauss}. Additionally, $\eta$
terms were not reported in \zhang\ as their bias contributions to cosmic shear were
found to be insignificant. Bearing in mind that differences in PSF treatment limit
interpretability, we compare our $\beta$ results; values reported for
HSC in \tabref{coefficients} have been adjusted by the ratio of
$\langle T_{\rm PSF}/T_{\rm gal}\rangle$ in the LSST-Y1 and HSC catalogs
($0.94/0.676$), which we justify by noting that the $\delta e^{(i)}$ moment-response
curves in \figref{mom-response} are (approximately) linear in
$T_{\rm PSF}/T_{\rm gal}$. The rescaled HSC-Y3 $\beta_2$ is consistent with
our result, given the HSC error bar; the $\beta_4$ value is consistent in magnitude
but has opposite sign to our result.

\section{Results}\label{sec:results}
We prepare the PSF residual catalogs based on the stellar simulations (described in
\secref{sims}) processed with \piff\ (see \secref{psf-modeling}), with PSFs
quantified using the moment-based parameters defined in \secref{moments}. 
The simulated visits cover our 100 square degree sky area to the expected number
of visits in $i$-band for the full 10-year LSST survey, so we refer to the full
catalog of residuals as our ``Y10'' PSF catalog. We use subsets of these visits
to construct ``Y1'' PSF catalogs---unless otherwise specified, the Y1 results we
quote are averaged over 50 catalogs constructed by sampling 150 visits, without
replacement, from the Y10 catalog (1508 visits total).

As described in \secref{measure}, we can examine PSF modeling with various metrics
of performance. In \secref{results-resid} we show and discuss ``one-point''
statistics of PSF residuals: one- and two-dimensional histograms of PSF parameters
and residuals, in \secref{results-rho} we focus on two-point statistics, and finally
in \secref{results-dxi} we propagate the PSF residual to a contamination on cosmic
shear and discuss implications.

\subsection{PSF parameter residuals}\label{sec:results-resid}
The distributions of residual PSF parameters are summarized by their mean and
standard deviations in \tabref{residual-stats}, given for the PSF (reserve) samples
in both Y1 and Y10 catalogs. We do not find evidence of overfitting, with the largest
difference in $\sigma$ between PSF and reserve stars being 7\% for
$\delta \tfour/\tfour$. The largest mean residual is also in $\delta \tfour / \tfour$
with a bias of $-6\times 10^{-3}$. Shape parameter residuals are all $< 2\times 10^{-5}$.

\begin{table}
    \centering
    \begin{tabular}{ccccc}
        \toprule
        & Y1  & Y10  & Y1 & Y10\\
        \cmidrule(r){2-3} \cmidrule(l){4-5}
        & \multicolumn{2}{c}{$\mu\times 10^{-3}$} & \multicolumn{2}{c}{$\sigma\times 10^{-2}$} \\
        \midrule
        $\delta \ttwo/\ttwo $ & 0.36 (0.37) & 0.33 (0.35) & 1.4 (1.4) & 1.4 (1.4) \\
        \addlinespace[.25em]
        $\delta \tfour/\tfour$ & $-$6.7 ($-$6.5) & $-$6 ($-$5.7) & 14 (15) & 14 (15) \\
        \addlinespace[.5em]
        & \multicolumn{2}{c}{$\mu \times 10^{-6}$} & \multicolumn{2}{c}{} \\
        \cmidrule(r){2-3}
        $\delta \gtwo_1$       & $-$18.3 ($-$10.7) & $-$18.1 ($-$7.9) & 1.8 (1.8) & 1.8 (1.8) \\
        \addlinespace[.25em]
        $\delta \gtwo_2$       & 2.7 (3.6) & 0.99 (4.5) & 1.8 (1.8) & 1.8 (1.8) \\
        \addlinespace[.25em]
        $\delta \efour_1$      & 12.7 (15.9) &  9.9 (17) & 1.4 (1.4) & 1.4 (1.4) \\
        \addlinespace[.25em]
        $\delta \efour_2$      & 4.5 ($-$6.3) & 4.9 ($-$2.8) & 1.4 (1.4) & 1.4 (1.4) \\
       \bottomrule
    \end{tabular}
    \caption{\label{tab:residual-stats}
        Mean and standard deviation of the PSF parameter residuals, for the PSF (reserve) stars in the Y1 and Y10 catalogs.
    }
\end{table}

For a quick estimate of the impact of these PSF size errors on multiplicative bias,
we use the \cite{hirata_ggl_2004} estimate for shear calibration uncertainty from errors in PSF
$\ttwo$, $\delta m = \langle T_{\rm PSF} / T_{\rm gal} \rangle \delta \ttwo / \ttwo$. 
With $\langle T_{\rm PSF} / T_{\rm gal} \rangle \sim 1$, we find $\delta m < 4 \times 10^{-4}$
for all mean residual $\ttwo$ in \tabref{residual-stats}, well under LSST DESC requirements
of $\delta m < 3\times 10^{-3}$ \citep{the_lsst_dark_energy_science_collaboration_lsst_2021}.
Gaussian assumptions prevent us from constructing an equivalent heuristic for the impact of
errors in PSF $\tfour$. This motivates future work extending the simulations developed in
\secref{coeff} to propagate higher-order PSF size contributions to multiplicative bias.

\begin{figure}
\includegraphics[width=0.47\textwidth]{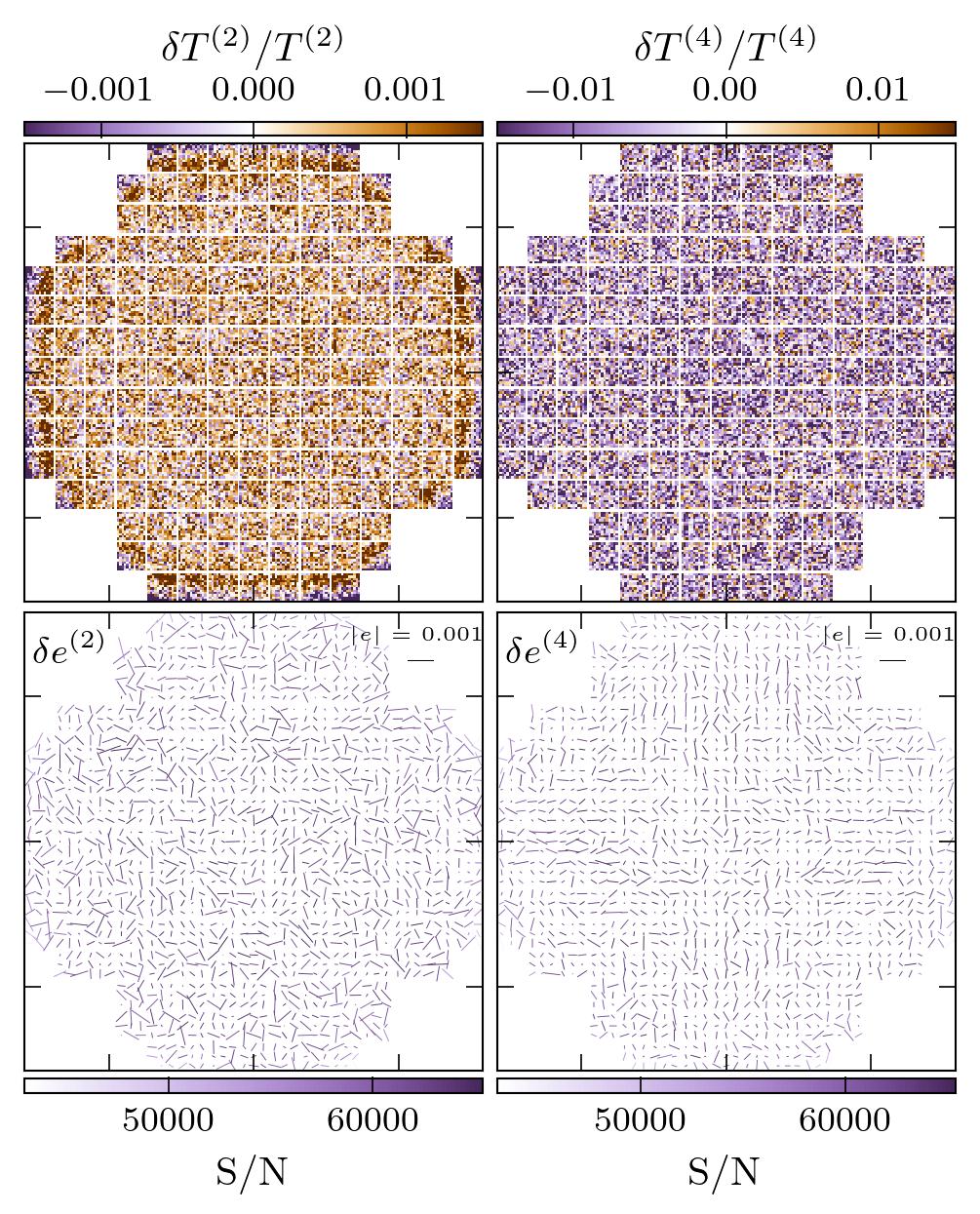}
\caption{\label{fig:residuals-fp}
    PSF parameter residuals stacked across all exposures as a function of focal plane position, 
    for reserve stars only. The upper left (right) panel shows the second- (fourth-)
    order scalar size residuals $\delta \ttwo / \ttwo$ ($\delta \tfour / \tfour$).
    The lower left (right) panel shows the second- (fourth-) order spin-2 shape
    residuals $\delta \gtwo$ ($\delta \efour$). The length and orientation of the
    whiskers show the magnitude and orientation of the residual PSF ellipse,
    and whisker color encodes the combined star S/N within each bin.
    }
\end{figure}

As mentioned in \secref{starsim}, the simulations are carried out with a model of
telescope optics and detectors, leading to differences in the PSF with location
on the focal plane. Although exposures in the LSST survey are dithered and rotated
to minimize the impact of such instrumental effects, it remains important to check for
residual patterns in focal plane coordinates to validate PSF model performance.
We show, in \figref{residuals-fp}, residuals for each PSF parameter in 181 LSSTCam
detectors, stacked over all 1508 exposures. The eight corner detectors are strongly
affected by vignetting, rendering sky subtraction very difficult. As they will a
priori not be included in the Rubin data releases, 
we do not include them in any part of this analysis. There is a ring pattern
visible on the outer edge of the focal plane in $\delta \ttwo/\ttwo$, likely
also due to vignetting. We see no evidence of other focal-plane level patterns
in PSF parameter residuals.

\begin{figure}
\includegraphics[width=0.47\textwidth]{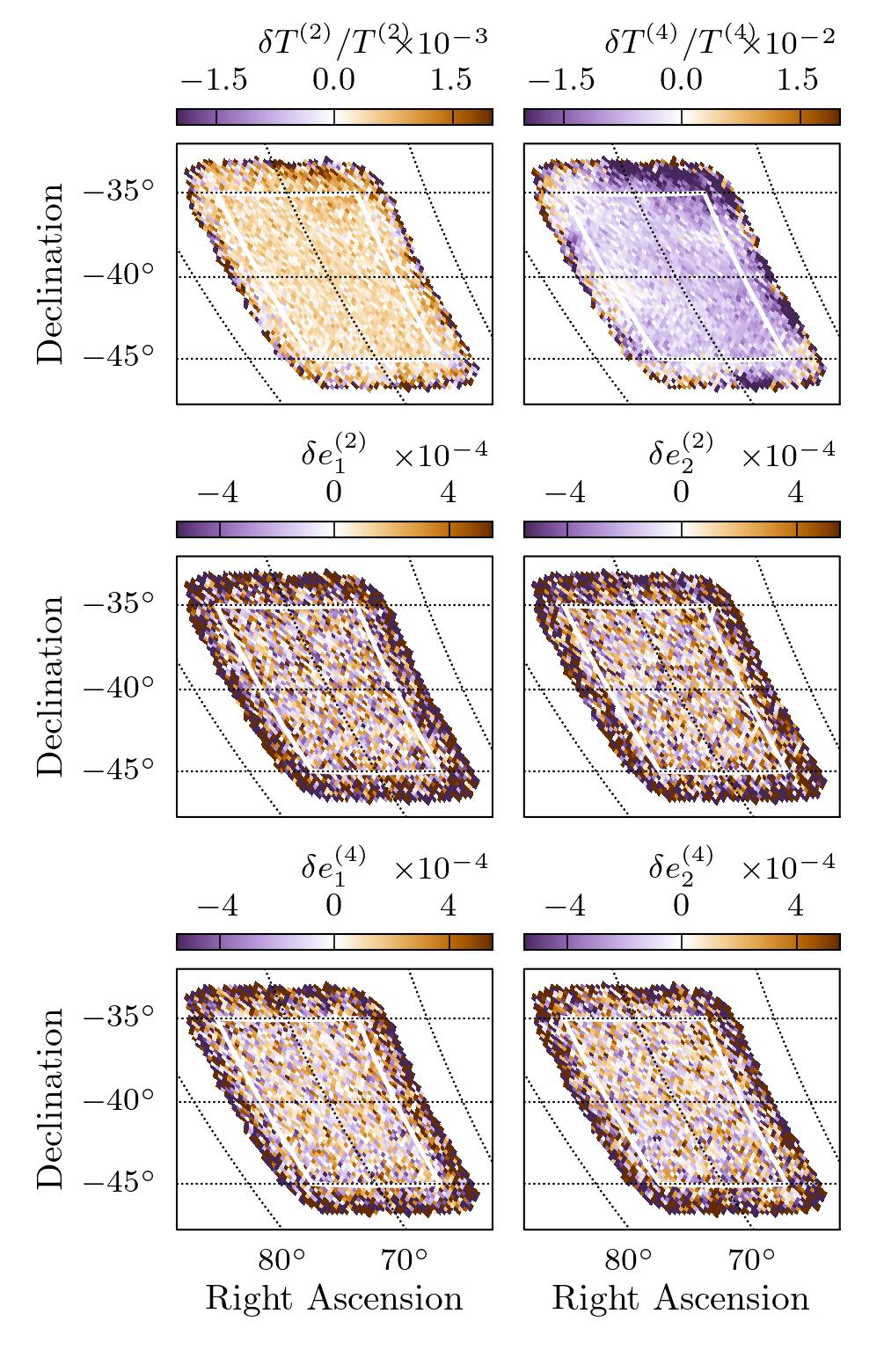}
\caption{\label{fig:residuals-sky}
    PSF parameter residuals as a function of position on sky. The white box denotes
    the 100 square degree full-depth area.
    }
\end{figure}

In \figref{residuals-sky} we show residuals from all 1508 visits in celestial
coordinates. The white box in each panel delineates the full-depth area; outside
the region we notice larger residuals as there are fewer exposures to average over
and a relatively larger amount of focal plane edges. The ring pattern seen in the
size residuals in \figref{residuals-fp} does not survive exposure dithering. Due
to the small sky area of our simulation we do not anticipate much spatial
variation in PSF model performance due to \eg\ stellar density or systematic
differences in observing conditions. Indeed, we do not see any particular residual
patterns in the residual sky maps other than the bias visible in $\delta \ttwo/\ttwo$
(mean positive, negligible at $4\times10^{-4}$) and $\delta \tfour/\tfour$ (mean
negative, magnitude $<10^{-2}$). 

Overall we find little cause for concern in the PSF residuals, so we continue
to two-point metrics in the following section. We do not discuss
color-dependence of residuals here since the $i$-band exposures are negligibly
impacted by chromatic atmospheric seeing, sensor, or airmass (DCR) effects.

\subsection{Measured two-point PSF statistics}\label{sec:results-rho}

\begin{figure*}
    \includegraphics[width=0.99\textwidth]{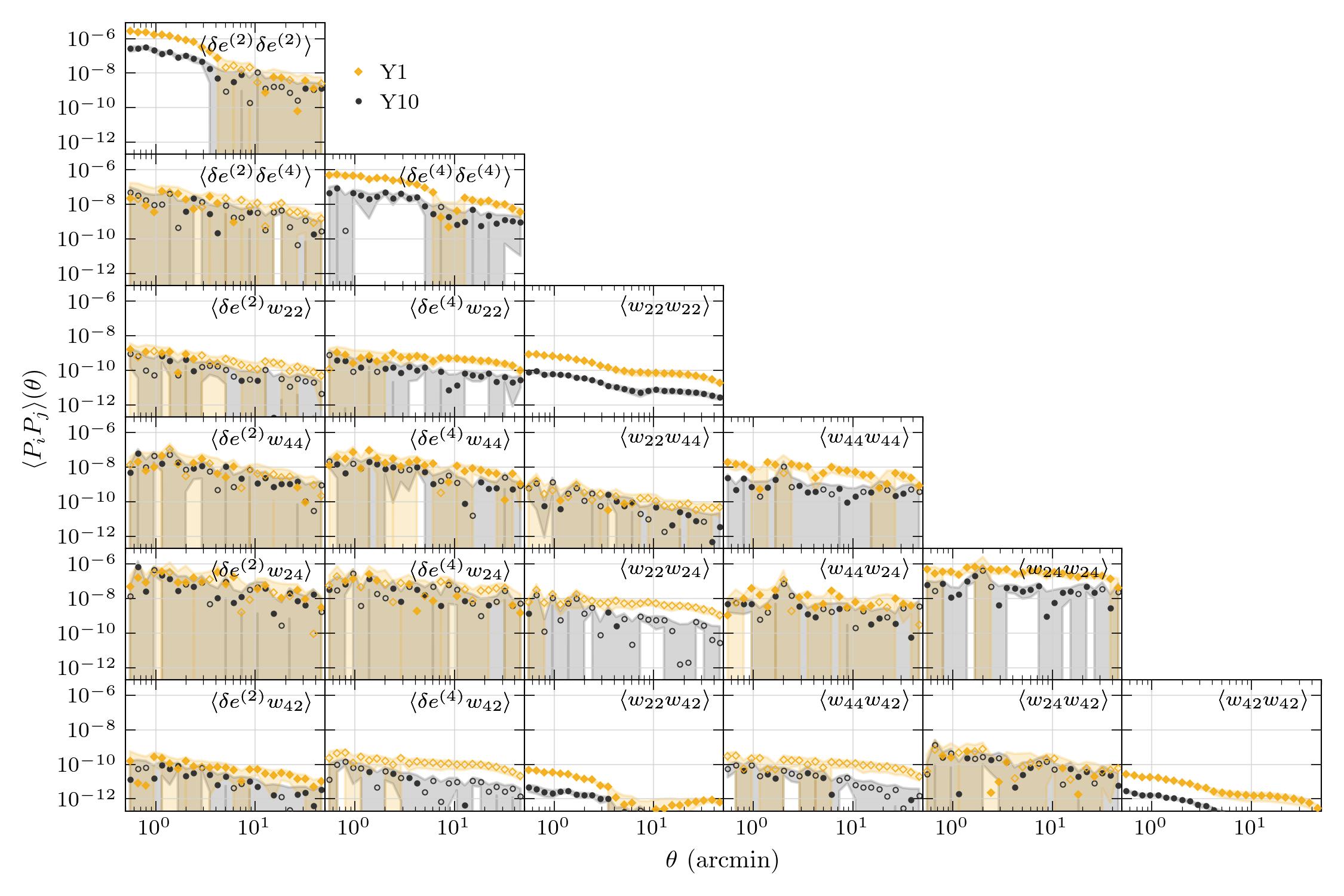}
    \caption{\label{fig:results-rho}
        The 21 rho statistics involving only PSF residual parameters (\ie, not $\gtwo$
        or $\efour$, those 15 functions are displayed in \figref{rho-shape}), for the $i$-band
        PSF simulation results of Y1 (gold diamonds) and Y10 (dark grey squares). Each column
        (row) has the same $i$th ($j$th) PSF parameter in the $\langle P_i P_j\rangle$
        correlation function. Shaded regions show error bars estimated from the spread
        over catalog results (gold, Y1), details in text, or from jackknife resampling (grey, Y10).
        }
    \end{figure*}

Measured correlations for the PSF residual-only rho statistics defined in
\secref{2pt-formalism} are shown in \figref{results-rho}. The 15 statistics
involving the PSF leakage term, which we assume will not contribute to
bias in cosmic shear since $\alpha\simeq0$ for \metadet and
\texttt{AnaCal}, are shown in \figref{rho-shape}. The correlations are calculated
with \texttt{TreeCorr}\footnote{\url{https://github.com/rmjarvis/TreeCorr}}
\citep{jarvis_skewness_2004}. The error bars on Y10 data points (grey bands) are
estimated with jackknife resampling. The errors on Y1 (gold bands) represent
the spread over correlation functions calculated from the 50 different Y1 catalogs,
divided by 10, the number of unique data splits possible.

Overall we find that our measured rho statistics have low amplitudes generally
indicative of good performance \citep[see \schutt, \zhang,][]{jefferson_reanalysis_2025};
we quantify this statement further in the next section.
The small sky area of our survey should affect the size of our error bars but not
the correlation function amplitude, which depends instead on the number of epochs
on average; therefore the amplitudes measured here correspond to what our
simulations predict for \textit{full-sky} $i$-band LSST data.

Specifically, we expect that the amplitude of a correlation function measured on
$N$ realizations of a random field should decrease as $1/N$. That we see a factor
of 10 between many of the Y1 (yellow) and Y10 (dark grey) points in \figref{results-rho},
at least for those correlations that are well-measured (\eg, $\langle w_{22}w_{22}\rangle$),
is a promising indication of independent residuals between exposures.

The rho statistics are calculated as a function of separation scale $\theta$, with
a maximum of 50\,\amin corresponding to roughly a quarter of the
size of the LSSTCam field. The shape residual autocorrelation functions
$\langle \delta e^{(i)}\delta e^{(j)}\rangle$ have the steepest scale dependence;
the decline in amplitude with $\theta$ indicates that, as expected, the second-order
polynomial used in the spatial interpolation is not able to completely capture the
atmospheric turbulence variation on small scales, but it successfully removes PSF
shape correlations on larger scales ($\theta\geq$ a few arcminutes).
The rho statistics involving the PSF size residual terms $w_{ij}$ in general show
a shallower decrease in correlation with increasing $\theta$ than the
$\langle \delta e^{(i)}\delta e^{(j)}\rangle$ statistics; however, this behavior
is harder to interpret since $w_{ij}$ entangles the fractional size residual
with the PSF shape $e^{(i)}$ (\eqnref{wij}).

\subsection{Predicted impact of PSF residuals on cosmic shear}\label{sec:results-dxi}

The measured rho statistics in \figref{results-rho} and the PSF contamination
coefficients in \tabref{coefficients} are the two ingredients we need to propagate
our PSF residual results to a predicted $\delta \xi^{\rm PSF}_{\pm}$ impact on cosmic
shear using \eqnref{dxi-psf}, which we repeat here:
\begin{equation*}
    \delta \xi_{\pm}^{\rm PSF} = \sum_{k=1}^{8}\sum_{l=1}^{8} c_k c_l \langle P_k P_l \rangle \,. 
\end{equation*}
The $i=2,j=2$ term with $\beta_2 \langle\delta\gtwo\delta\gtwo\rangle$ is, at small $\theta$, the most significant contribution to $\delta\xi_+^{\rm PSF}$ (see \tabref{coefficients} and \figref{results-rho}), but other terms also contribute---to find the full impact on cosmic shear we combine them all.

\begin{figure}
    \includegraphics[width=0.47\textwidth]{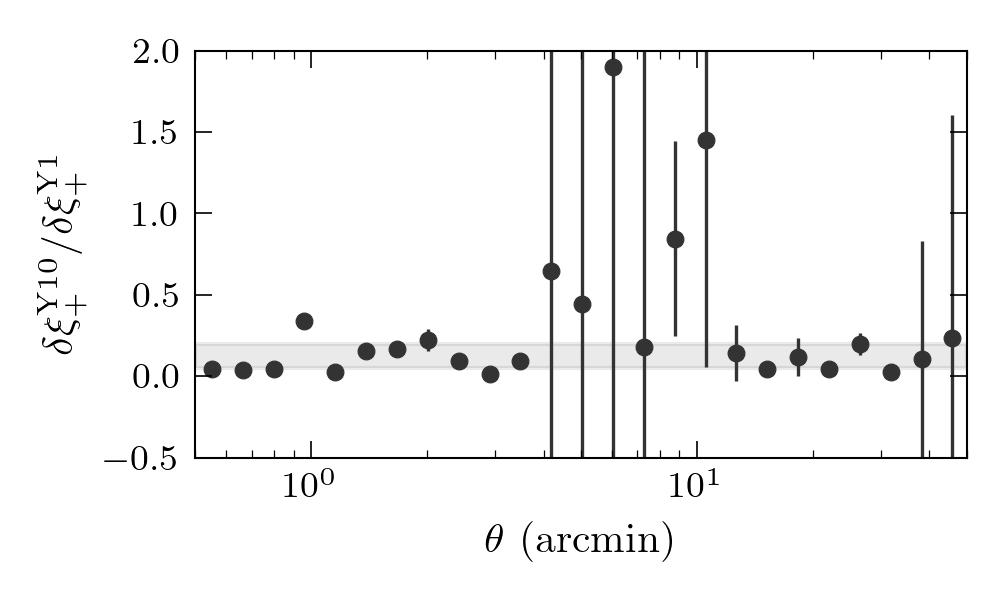}
    \caption{\label{fig:dxi-ratio}
        Ratio of $i$-band $\delta \xi^{\rm PSF}_+$ for Y10 to Y1 (dark grey circles),
        with error bars bootstrapped over the Y1 samples. The shaded region lies between
        $2\times$ and $0.5\times$ the expected $N^{Y1}_{\rm epoch}/N^{Y10}_{\rm epoch}=0.1$ result.
    }
    \end{figure}

Our desired result is a full multi-band $r+i+z$ estimate of $\delta \xi^{\rm PSF}_+$, but we have only $i$-band simulations. In \schutt\ (see their figures 20 and 24) the $i$-band rho statistics are shown to lie between the $r$- and $z$- band at
large separations, and are not significantly different at small separations.
Therefore, our results from $i$-band simulations may represent an effective
average of combined $riz$ statistics. We must still account for the expected
increase in total number of epochs that would result from including the $r$
and $z$-band data, since, as discussed in \secref{results-rho}, the rho
statistics—--and thus their linear combination in $\delta \xi^{\rm PSF}_+$---are
expected to scale as $1/N_{\rm epoch}$. We show in \figref{dxi-ratio} that the ratio of
$\delta \xi^{\rm PSF}_+$ for Y10 and Y1 is consistent with this scaling, so
in what follows we use $N_{\rm epoch}^{i}/N_{\rm epoch}^{riz}\simeq0.35$
(see \secref{galaxysim}) to scale our $i$-band $\delta \xi_+^{\rm PSF}$ results
to a $riz$ multi-band prediction.

\begin{figure}
    \includegraphics[width=0.47\textwidth]{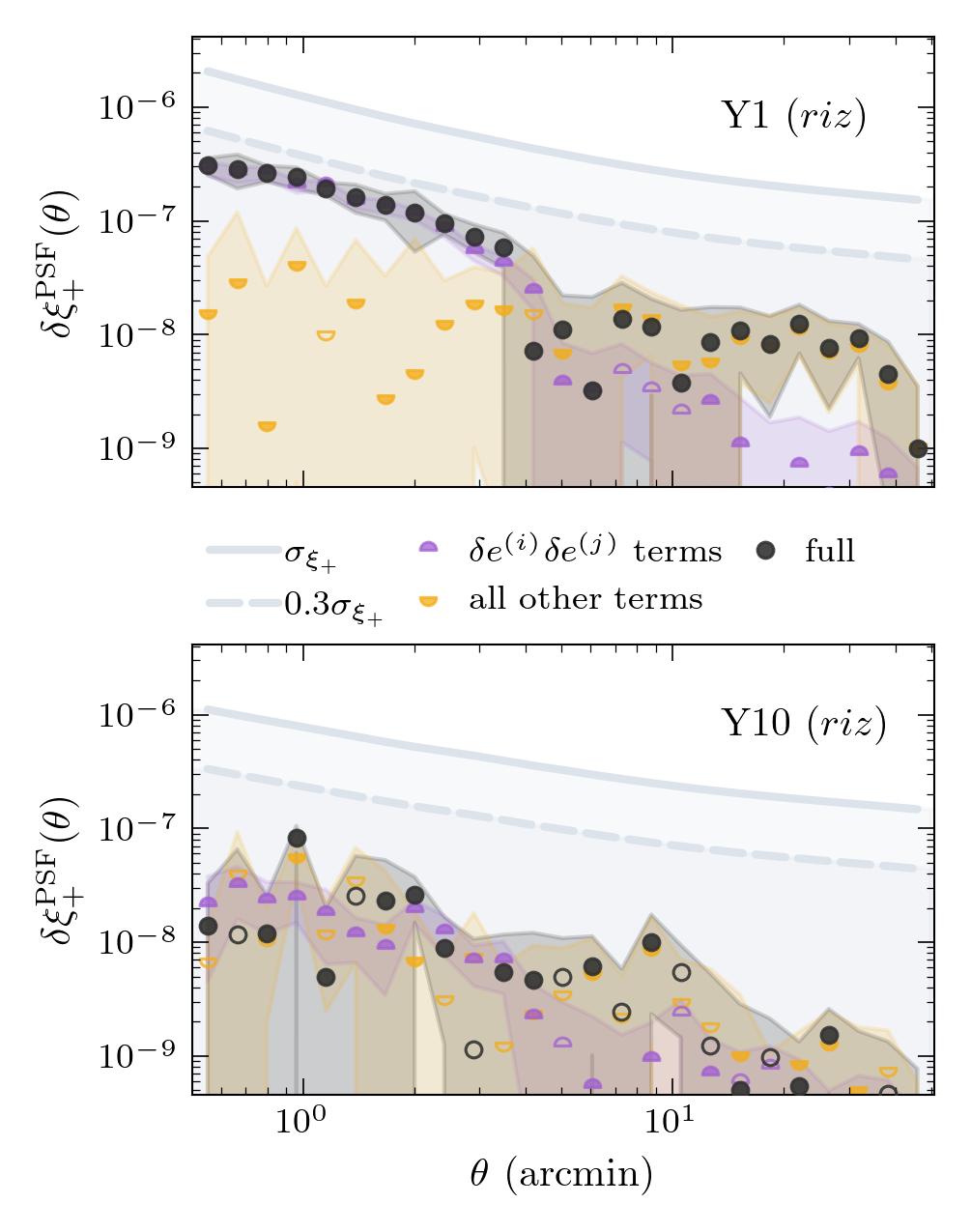}
    \caption{\label{fig:dxi}
        Multi-band $riz$ estimate of $\delta \xi_+^{\rm PSF}$ (dark grey circles),
        the PSF contamination to cosmic shear, for Y1 (upper) and Y10 (lower).
        The contributions to $\delta \xi_+^{\rm PSF}$ from only shear-residual
        terms (purple upper half-circles) and from all other terms (gold lower
        half-circles) are also shown. Colored (dark grey, purple, gold) shaded
        regions show the 1\,$\sigma$ errors on the PSF contamination data points.
        The light grey lines correspond to the 1\,$\sigma_{\xi_+}$
        (solid line) and 0.3\,$\sigma_{\xi_+}$ (dashed line) cosmic shear uncertainty estimated from our analytical covariance matrix.
        }
    \end{figure}

These final results are shown in \figref{dxi}. The cosmic shear uncertainty
$\sigma_{\xi_+}$ (see \secref{cov}) for Y1 (upper panel) and Y10 (lower panel)
is plotted in light grey with solid lines. Ideally the PSF contamination will
contribute a bias significantly lower than the full cosmic shear uncertainty,
so we show a more stringent ``requirement'' of 30\% (light grey dashed lines) as
well. The multi-band $\delta \xi_+^{\rm PSF}$ is shown in black points; we find
that at all scales $\theta$ for both Y1 and Y10, the PSF contamination is below
0.3\,$\sigma_{\xi_+}$. At scales $\theta\geq5\amin$ most relevant for cosmic shear
analyses (\ie, likely to be included after scale cuts), the contamination is
\textit{well} below this level.
We measure, but do not detect, $\xi_-$; the noise level, an upper bound for the $\xi_-$ signal, is well below $\sigma_{\xi_+}$.

We can determine how particular rho statistics are contributing to the full contamination
result by splitting the sum in \eqnref{dxi-psf} into multiple terms; we show in \figref{dxi}
the contribution from only $\langle \delta e^{(i)}\delta e^{(j)}\rangle$ terms in purple
markers, and from all other terms in gold---\ie, the dark grey data points are the sum of
the purple and the gold. In the Y1 results, these two contributions to $\delta \xi_+^{\rm PSF}$
show behavior similar to what was observed in the rho statistics in \figref{results-rho},
with pure shape residuals dominating the PSF contamination at $\theta < 5\,\amin$ but
becoming negligible compared to the contribution from all other terms at larger scales.
At these scales $\theta \geq 5\,\amin$, the Y1 $\delta \xi_+^{\rm PSF}$ is fairly flat, behavior
consistent with the rho statistics involving $w_{ij}$.
The Y10 result for the shape-residual contribution is roughly $10\times$ smaller on all scales
than its Y1 counterpart. The Y10 contribution from all other terms behaves differently than
in Y1, however: at small scales it is of similar amplitude, but measured more robustly, while
at larger scales the correlation is noisier but drops off more steeply with $\theta$.

\section{Conclusions and Future Work}\label{sec:conclusions}

In this work we present an evaluation on simulated data of PSF modeling performance
for cosmic shear measurements with LSST. We find that, given the effects included
in our semi-realistic simulated images, the \piff\ modeling package performs quite
well. In particular, the bias from PSF contamination to two-point correlations
functions of galaxy shear falls well within acceptable limits for cosmology studies,
for both Y1 and Y10 LSST data.

Importantly, however, we did not include certain effects into our simulations which are
known to cause difficulties in PSF modeling; we detail a few significant examples below.
In particular, some of the omitted effects may result in PSF modeling errors that
are correlated across exposures; in this case, the improvement from Y1 to Y10 would in 
reality be reduced relative to what is implied by the simulation results.
\begin{enumerate}
\item We did not include detector-level variation between sensors in the focal plane
or pointing-related variations in the optical PSF, both of which are still being
characterized during LSSTCam and Rubin commissioning but have the potential to make
the spatial interpolation of the PSF more challenging \citep{rubinobs_dp1_2025}.
\item We assumed the CCD surfaces are flat, which is known not to be perfectly true.  For slightly out-of-focus images, sensor height variation could impact the PSF size. 
\item We assumed the WCS was perfectly known and accounted for correctly.  If there are astrometric errors due to electric field effects, which are not fully included in the WCS, these would also impact the PSF size and shape.
\item We have minimized the effects of chromatic seeing
and differential chromatic refraction (DCR) by simulating only in $i$-band, where these
effects are smallest, and ignored any SED differences between the stars and galaxies in
our propagation from PSF errors to galaxy shape errors via the PSF contamination
coefficients described in \secref{coeff}. Corrections for these effects exist at levels
sufficient for DES-Y6 (\schutt) but are still in the research and development stage to
meet Rubin cosmic shear requirements \citep{meyers_impact_2015}.
\end{enumerate}

These are important contributors to the PSF modeling performance, but as characterization
and method development to model these effects at the level of Rubin requirements is
ongoing, we leave their evaluation to future work. In the meantime, we conclude that
other aspects of the PSF modeling in the \lsstpipe, such as removing persistent patterns
in the atmospheric PSF, already perform quite well on our simulations. Real data from
Rubin will soon enable more in-depth analyses. 

Another subtlety in characterizing PSF performance for Rubin is that shear will be
measured on cell-based coadded images and using a coadded PSF \citep{armstrong_little_2024,
mandelbaum_psfs_2023}. In this scheme, shear within a small region (cell) is measured
using the co-add PSF at the center of the cell; this will introduce some additional error
away from the center. \cite{sheldon_metadetection_2023}, using the perfect PSF model for
the center of the cells, found this away-from-center error to be insignificant. Our
analysis, on the other hand, was performed on single-exposure PSFs to isolate errors due
to PSF modeling. The effect on shear bias of cell-based coadd measurements with imperfect
PSF models, or with imperfect astrometry due to atmospheric turbulence, remains to be studied on real data. 

\section*{Acknowledgements}
This paper has undergone internal review in the LSST Dark Energy Science Collaboration. The internal reviewers were Axel Guinot, Mike Jarvis, and Pierre-Fran\c{c}ois L\'eget.
We thank Xiangchong Li, Theo Schutt, and Masaya Yamamoto for fruitful discussions.
We thank the excellent computing staffs of the Scientific Data and Computing Center (SDCC) at Brookhaven National Laboratory.
The DESC acknowledges ongoing support from the Institut National de Physique Nucl\'eaire et de Physique des Particules in France; the Science \& Technology Facilities Council in the United Kingdom; and the Department of Energy and the LSST Discovery Alliance in the United States.
DESC uses resources of the IN2P3 Computing Center (CC-IN2P3--Lyon/Villeurbanne - France) funded by the Centre National de la Recherche Scientifique; the National Energy Research Scientific Computing Center, a DOE Office of Science User Facility supported by the Office of Science of the U.S.\ Department of Energy under Contract No.\ DE-AC02-05CH11231; STFC DiRAC HPC Facilities, funded by UK BEIS National E-infrastructure capital grants; and the UK particle physics grid, supported by the GridPP Collaboration.
This work was performed in part under DOE Contract DE-AC02-76SF00515.

This paper makes use of LSST Science Pipelines software developed by the Vera C. Rubin Observatory. We thank the Rubin Observatory for making their code available as free software at \url{https://pipelines.lsst.io}.

\section*{Author Contributions}
CAH: defined the project, contributed to the \imsim pipeline, developed and ran the postage stamp simulations of shear response, performed the end-to-end data analysis, and wrote the paper. 
ESS: contributed to defining the project and advised throughout, wrote a library to interface with \imsim, ran \imsim and processed the images with \piff, and generated the galaxy catalog.
TZ: contributed code and expertise with higher-order moments to the shear response simulations. JHD: ran covariance calculations. MJ: Internal Reviewer, DESC Builder; helped develop \piff software used in this paper. All authors contributed to the paper text.

\appendix \label{app:rho-e}
\renewcommand{\thefigure}{A.\arabic{figure}}
\setcounter{figure}{0}

\section*{Impact of non-zero PSF leakage}
In the main text we assumed that the PSF leakage coefficients $\alpha_2, \alpha_4=0$.
This is a well-motivated assumption, since the DES-Y6-measured values (see
\tabref{coefficients}) are consistent with zero as expected for \metadet.
Here, we explore how our conclusions might change under a different assumption,
namely $\alpha_2=0.01$, $\alpha_4=-0.01$. The sign difference is motivated by the
negative correlation found between these coefficients in \cite{yamamoto_dark_2025}.

\begin{figure*}
    \includegraphics[width=0.99\textwidth]{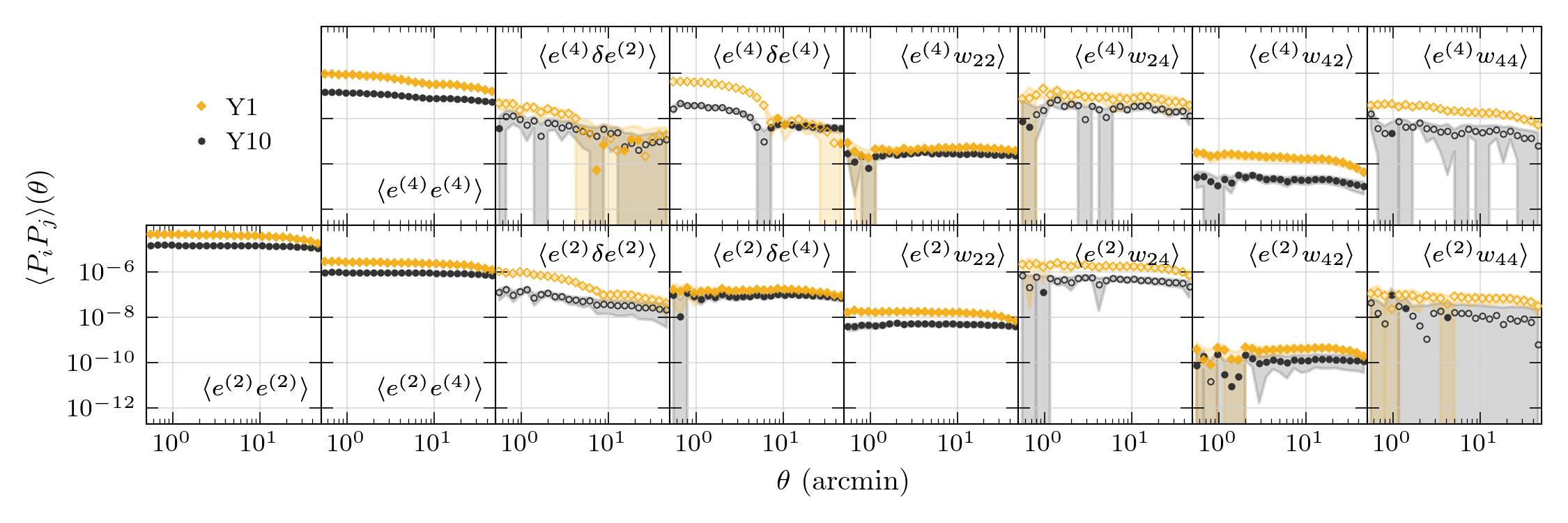}
    \caption{\label{fig:rho-shape}
        The 15 rho statistics involving $\gtwo$ or $\efour$ shape directly,
        for the $i$-band PSF simulation results of
        Y1 (gold diamonds) and Y10 (dark grey squares). Each row (column) has
        the same $i$th ($j$th) PSF parameter in the $\langle P_i P_j\rangle$
        correlation function. Shaded regions show error bars estimated from the
        spread over catalog results (gold, Y1), details in text, or from
        jackknife resampling (grey, Y10).
        }
    \end{figure*}

First, we show the rho statistics involving PSF shapes $\gtwo$, $\efour$ in
\figref{rho-shape}. Most of these measured correlations show a shallower drop-off
in both $N_{\rm epoch}$, going from Y1 to Y10, and in scale $\theta$, than the rho
stats of PSF residual parameters. These trends indicate the presence of
correlations on large spatial scales which persist between exposures, \eg, the
atmospheric patterns in PSF ellipticity caused by the prevailing wind direction
at the Rubin site expected in our simulations \citep{hebert_generation_2024}. 
As discussed in the main text, these patterns seem to be well-modeled
by \piff since we did not see evidence of any preferred orientation in the PSF
parameter residuals. In the (unexpected) presence of PSF leakage, however,
persistent patterns in PSF ellipticity could still cause bias to the shape measurements.

\begin{figure}
    \includegraphics[width=0.47\textwidth]{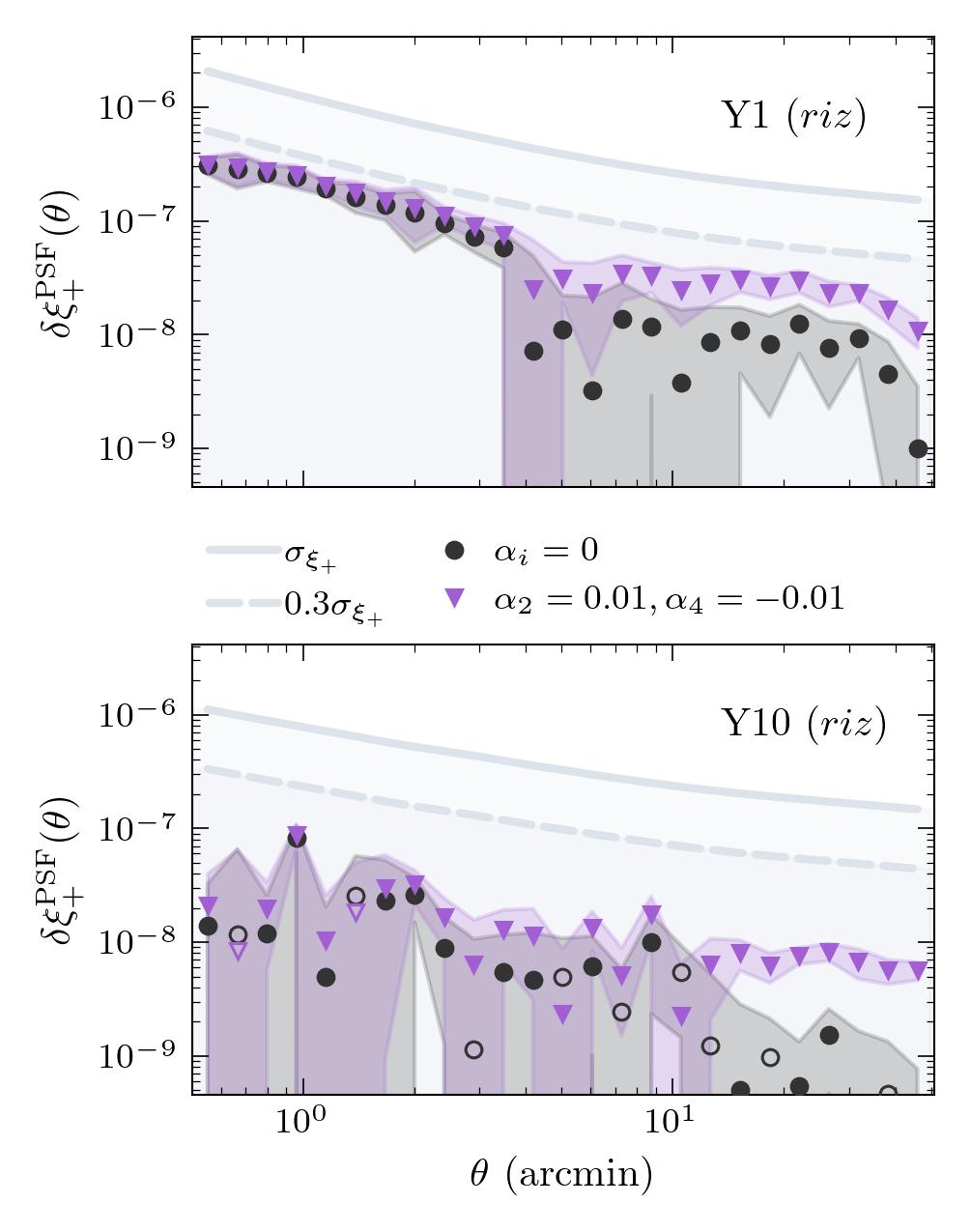}
    \caption{\label{fig:dxi-alpha-nonzero}
        Multi-band $riz$ estimates of $\delta \xi_+^{\rm PSF}$, the PSF contamination
        to cosmic shear, for Y1 (upper) and Y10 (lower). Fiducial result shown in dark grey
        circles, and results with $\alpha_2, \alpha4 \neq 0$ overlaid in purple triangles.
        The light grey lines correspond to the 1\,$\sigma_{\xi_+}$
        (solid line) and 0.3\,$\sigma_{\xi_+}$ (dashed line) cosmic shear uncertainty estimated from our analytical covariance matrix.
        }
    \end{figure}

In \figref{dxi-alpha-nonzero} we show results for the leakage-added PSF contamination
to cosmic shear. Since \figref{rho-shape} shows that many of the rho statistics
involving PSF shape do not scale as $1/N_{\rm epoch}$, we scaled only the non-PSF-leakage
terms by the $riz$ exposure numbers, but did not alter the PSF leakage
contributions from their $i$-band values. This will give us a slightly pessimistic
estimate. The results in \figref{dxi-alpha-nonzero} indicate a larger PSF
contamination than without PSF leakage, as expected, but the increase in
$\delta \xi_+^{\rm PSF}$ seems to mostly affect the larger scales, and even with
the pessimistic estimate of the contamination we are still comfortably under 30\%
of the cosmic shear uncertainty $\sigma_{\xi_+}$.

\bibliography{bibliography}

@article{jarvis_skewness_2004,
	title = {The {Skewness} of the {Aperture} {Mass} {Statistic}},
	volume = {352},
	issn = {00358711, 13652966},
	url = {http://arxiv.org/abs/astro-ph/0307393},
	doi = {10.1111/j.1365-2966.2004.07926.x},
	number = {1},
	journal = {Monthly Notices of the Royal Astronomical Society},
	author = {Jarvis, M. and Bernstein, G. and Jain, B.},
	month = jul,
	year = {2004},
	note = {arXiv:astro-ph/0307393},
	pages = {338--352}
}

@article{zhang_general_2023,
	title = {A {General} {Framework} for {Removing} {Point} {Spread} {Function} {Additive} {Systematics} in {Cosmological} {Weak} {Lensing} {Analysis}},
	issn = {0035-8711, 1365-2966},
	url = {http://arxiv.org/abs/2212.03257},
	doi = {10.1093/mnras/stad1801},
	journal = {Monthly Notices of the Royal Astronomical Society},
	author = {Zhang, Tianqing and Li, Xiangchong and Dalal, Roohi and Mandelbaum, Rachel and Strauss, Michael A. and Kannawadi, Arun and Miyatake, Hironao and Nicola, Andrina and Malagón, Andrés A. Plazas and Shirasaki, Masato and Sugiyama, Sunao and Takada, Masahiro},
	month = jun,
	year = {2023},
	note = {arXiv:2212.03257 [astro-ph]},
	pages = {stad1801}
}

@article{zhang_impact_2021,
	title = {Impact of point spread function higher moments error on weak gravitational lensing},
	volume = {510},
	issn = {0035-8711, 1365-2966},
	url = {https://academic.oup.com/mnras/article/510/2/1978/6460504},
	doi = {10.1093/mnras/stab3584},
	number = {2},
	journal = {Monthly Notices of the Royal Astronomical Society},
	author = {Zhang, Tianqing and Mandelbaum, Rachel and {the LSST Dark Energy Science Collaboration} and Kannawadi, Arun and Miyatake, Hironao and Astier, Pierre and Jarvis, Mike and Meyers, Josh and Schmitz, Morgan and Clowe, Douglas},
	month = dec,
	year = {2021},
	pages = {1978--1993}
}

@article{zhang_impact_2023,
	title = {Impact of point spread function higher moments error on weak gravitational lensing – {II}. {A} comprehensive study},
	volume = {520},
	issn = {0035-8711, 1365-2966},
	url = {https://academic.oup.com/mnras/article/520/2/2328/6835521},
	doi = {10.1093/mnras/stac3350},
	number = {2},
	journal = {Monthly Notices of the Royal Astronomical Society},
	author = {Zhang, Tianqing and Almoubayyed, Husni and Mandelbaum, Rachel and Meyers, Joshua E and Jarvis, Mike and Kannawadi, Arun and Schmitz, Morgan A and Guinot, Axel and {the LSST Dark Energy Science Collaboration}},
	month = feb,
	year = {2023},
	pages = {2328--2350},
}

@article{sheldon_metadetection_2023,
	title = {Metadetection {Weak} {Lensing} for the {Vera} {C}. {Rubin} {Observatory}},
	volume = {6},
	issn = {2565-6120},
	url = {https://astro.theoj.org/article/75280-metadetection-weak-lensing-for-the-vera-c-rubin-observatory},
	doi = {10.21105/astro.2303.03947},
	journal = {The Open Journal of Astrophysics},
	author = {Sheldon, Erin S. and Becker, Matthew R. and Jarvis, Michael and Armstrong, Robert and {the LSST Dark Energy Science Collaboration}},
	month = may,
	year = {2023},
	pages = {10.21105/astro.2303.03947},
}

@article{the_lsst_dark_energy_science_collaboration_lsst_2021,
	title = {The {LSST} {DESC} {DC2} {Simulated} {Sky} {Survey}},
	volume = {253},
	issn = {0067-0049},
	url = {https://dx.doi.org/10.3847/1538-4365/abd62c},
	doi = {10.3847/1538-4365/abd62c},
	number = {1},
	journal = {The Astrophysical Journal Supplement Series},
	author = {{LSST Dark Energy Science Collaboration} and Abolfathi, Bela and Alonso, David and Armstrong, Robert and Aubourg, Éric and Awan, Humna and Babuji, Yadu N. and Bauer, Franz Erik and Bean, Rachel and Beckett, George and Biswas, Rahul and Bogart, Joanne R. and Boutigny, Dominique and Chard, Kyle and Chiang, James and Claver, Chuck F. and Cohen-Tanugi, Johann and Combet, Céline and Connolly, Andrew J. and Daniel, Scott F. and Digel, Seth W. and Drlica-Wagner, Alex and Dubois, Richard and Gangler, Emmanuel and Gawiser, Eric and Glanzman, Thomas and Gris, Phillipe and Habib, Salman and Hearin, Andrew P. and Heitmann, Katrin and Hernandez, Fabio and Hložek, Renée and Hollowed, Joseph and Ishak, Mustapha and Ivezi\'c, \v{Z}eljko and Jarvis, Mike and Jha, Saurabh W. and Kahn, Steven M. and Kalmbach, J. Bryce and Kelly, Heather M. and Kovacs, Eve and Korytov, Danila and Krughoff, K. Simon and Lage, Craig S. and Lanusse, François and Larsen, Patricia and Le Guillou, Laurent and Li, Nan and Longley, Emily Phillips and Lupton, Robert H. and Mandelbaum, Rachel and Mao, Yao-Yuan and Marshall, Phil and Meyers, Joshua E. and Moniez, Marc and Morrison, Christopher B. and Nomerotski, Andrei and O’Connor, Paul and Park, HyeYun and Park, Ji Won and Peloton, Julien and Perrefort, Daniel and Perry, James and Plaszczynski, Stéphane and Pope, Adrian and Rasmussen, Andrew and Reil, Kevin and Roodman, Aaron J. and Rykoff, Eli S. and Sánchez, F. Javier and Schmidt, Samuel J. and Scolnic, Daniel and Stubbs, Christopher W. and Tyson, J. Anthony and Uram, Thomas D. and Villarreal, Antonio and Walter, Christopher W. and Wiesner, Matthew P. and Wood-Vasey, W. Michael and Zuntz, Joe},
	month = mar,
	year = {2021},
	note = {Publisher: The American Astronomical Society},
	pages = {31},
}

@article{jarvis_science_2016,
	title = {The {DES} {Science} {Verification} weak lensing shear catalogues},
	volume = {460},
	issn = {0035-8711, 1365-2966},
	url = {https://academic.oup.com/mnras/article-lookup/doi/10.1093/mnras/stw990},
	doi = {10.1093/mnras/stw990},
	number = {2},
	journal = {Monthly Notices of the Royal Astronomical Society},
	author = {Jarvis, M. and Sheldon, E. and Zuntz, J. and Kacprzak, T. and Bridle, S. L. and Amara, A. and Armstrong, R. and Becker, M. R. and Bernstein, G. M. and Bonnett, C. and Chang, C. and Das, R. and Dietrich, J. P. and Drlica-Wagner, A. and Eifler, T. F. and Gangkofner, C. and Gruen, D. and Hirsch, M. and Huff, E. M. and Jain, B. and Kent, S. and Kirk, D. and MacCrann, N. and Melchior, P. and Plazas, A. A. and Refregier, A. and Rowe, B. and Rykoff, E. S. and Samuroff, S. and Sánchez, C. and Suchyta, E. and Troxel, M. A. and Vikram, V. and Abbott, T. and Abdalla, F. B. and Allam, S. and Annis, J. and Benoit-Lévy, A. and Bertin, E. and Brooks, D. and Buckley-Geer, E. and Burke, D. L. and Capozzi, D. and Carnero Rosell, A. and Carrasco Kind, M. and Carretero, J. and Castander, F. J. and Clampitt, J. and Crocce, M. and Cunha, C. E. and D'Andrea, C. B. and da Costa, L. N. and DePoy, D. L. and Desai, S. and Diehl, H. T. and Doel, P. and Fausti Neto, A. and Flaugher, B. and Fosalba, P. and Frieman, J. and Gaztanaga, E. and Gerdes, D. W. and Gruendl, R. A. and Gutierrez, G. and Honscheid, K. and James, D. J. and Kuehn, K. and Kuropatkin, N. and Lahav, O. and Li, T. S. and Lima, M. and March, M. and Martini, P. and Miquel, R. and Mohr, J. J. and Neilsen, E. and Nord, B. and Ogando, R. and Reil, K. and Romer, A. K. and Roodman, A. and Sako, M. and Sanchez, E. and Scarpine, V. and Schubnell, M. and Sevilla-Noarbe, I. and Smith, R. C. and Soares-Santos, M. and Sobreira, F. and Swanson, M. E. C. and Tarle, G. and Thaler, J. and Thomas, D. and Walker, A. R. and Wechsler, R. H.},
	month = aug,
	year = {2016},
	pages = {2245--2281},
}

@article{gatti_dark_2021,
	title = {Dark {Energy} {Survey} {Year} 3 {Results}: {Weak} {Lensing} {Shape} {Catalogue}},
	volume = {504},
	issn = {0035-8711, 1365-2966},
	shorttitle = {Dark {Energy} {Survey} {Year} 3 {Results}},
	url = {http://arxiv.org/abs/2011.03408},
	doi = {10.1093/mnras/stab918},
	number = {3},
	journal = {Monthly Notices of the Royal Astronomical Society},
	author = {Gatti, M. and Sheldon, E. and Amon, A. and Becker, M. and Troxel, M. and Choi, A. and Doux, C. and MacCrann, N. and Alsina, A. Navarro and Harrison, I. and Gruen, D. and Bernstein, G. and Jarvis, M. and Secco, L. F. and Ferté, A. and Shin, T. and McCullough, J. and Rollins, R. P. and Chen, R. and Chang, C. and Pandey, S. and Tutusaus, I. and Prat, J. and Elvin-Poole, J. and Sanchez, C. and Plazas, A. A. and Roodman, A. and Zuntz, J. and Abbott, T. M. C. and Aguena, M. and Allam, S. and Annis, J. and Avila, S. and Bacon, D. and Bertin, E. and Bhargava, S. and Brooks, D. and Burke, D. L. and Rosell, A. Carnero and Kind, M. Carrasco and Carretero, J. and Castander, F. J. and Conselice, C. and Costanzi, M. and da Costa, L. N. and Davis, T. M. and De Vicente, J. and Desai, S. and Diehl, H. T. and Dietrich, J. P. and Doel, P. and Drlica-Wagner, A. and Eckert, K. and Everett, S. and Ferrero, I. and Frieman, J. and García-Bellido, J. and Gerdes, D. W. and Giannantonio, T. and Gruendl, R. A. and Gschwend, J. and Gutierrez, G. and Hartley, W. G. and Hinton, S. R. and Hollowood, D. L. and Honscheid, K. and Hoyle, B. and Huff, E. M. and Huterer, D. and Jain, B. and James, D. J. and Jeltema, T. and Krause, E. and Kron, R. and Kuropatkin, N. and Lima, M. and Maia, M. A. G. and Marshall, J. L. and Miquel, R. and Morgan, R. and Myles, J. and Palmese, A. and Paz-Chinchón, F. and Rykoff, E. S. and Samuroff, S. and Sanchez, E. and Scarpine, V. and Schubnell, M. and Serrano, S. and Sevilla-Noarbe, I. and Smith, M. and Suchyta, E. and Swanson, M. E. C. and Tarle, G. and Thomas, D. and To, C. and Tucker, D. L. and Varga, T. N. and Wechsler, R. H. and Weller, J. and Wester, W. and Wilkinson, R. D.},
	month = may,
	year = {2021},
	note = {arXiv: 2011.03408},
	pages = {4312--4336},
}

@article{jarvis_dark_2020,
	title = {Dark {Energy} {Survey} {Year} 3 {Results}: {Point}-{Spread} {Function} {Modeling}},
	volume = {501},
	issn = {0035-8711, 1365-2966},
	shorttitle = {Dark {Energy} {Survey} {Year} 3 {Results}},
	url = {http://arxiv.org/abs/2011.03409},
	doi = {10.1093/mnras/staa3679},
	number = {1},
	journal = {Monthly Notices of the Royal Astronomical Society},
	author = {Jarvis, M. and Bernstein, G. M. and Amon, A. and Davis, C. and Léget, P. F. and Bechtol, K. and Harrison, I. and Gatti, M. and Roodman, A. and Chang, C. and Chen, R. and Choi, A. and Desai, S. and Drlica-Wagner, A. and Gruen, D. and Gruendl, R. A. and Hernandez, A. and MacCrann, N. and Meyers, J. and Navarro-Alsina, A. and Pandey, S. and Plazas, A. A. and Secco, L. F. and Sheldon, E. and Troxel, M. A. and Vorperian, S. and Wei, K. and Zuntz, J. and Abbott, T. M. C. and Aguena, M. and Allam, S. and Avila, S. and Bhargava, S. and Bridle, S. L. and Brooks, D. and Rosell, A. Carnero and Kind, M. Carrasco and Carretero, J. and Costanzi, M. and da Costa, L. N. and De Vicente, J. and Diehl, H. T. and Doel, P. and Everett, S. and Flaugher, B. and Fosalba, P. and Frieman, J. and García-Bellido, J. and Gaztanaga, E. and Gerdes, D. W. and Gutierrez, G. and Hinton, S. R. and Hollowood, D. L. and Honscheid, K. and James, D. J. and Kent, S. and Kuehn, K. and Kuropatkin, N. and Lahav, O. and Maia, M. A. G. and March, M. and Marshall, J. L. and Melchior, P. and Menanteau, F. and Miquel, R. and Ogando, R. L. C. and Paz-Chinchón, F. and Rykoff, E. S. and Sanchez, E. and Scarpine, V. and Schubnell, M. and Serrano, S. and Sevilla-Noarbe, I. and Smith, M. and Suchyta, E. and Swanson, M. E. C. and Tarle, G. and Varga, T. N. and Walker, A. R. and Wester, W. and Wilkinson, R. D. and Collaboration, D. E. S.},
	month = dec,
	year = {2020},
	note = {arXiv: 2011.03409},
	pages = {1282--1299},
}

@article{meyers_impact_2015,
	title = {{IMPACT} {OF} {ATMOSPHERIC} {CHROMATIC} {EFFECTS} {ON} {WEAK} {LENSING} {MEASUREMENTS}},
	volume = {807},
	issn = {1538-4357},
	url = {https://iopscience.iop.org/article/10.1088/0004-637X/807/2/182},
	doi = {10.1088/0004-637X/807/2/182},
	number = {2},
	journal = {The Astrophysical Journal},
	author = {Meyers, Joshua E. and Burchat, Patricia R.},
	month = jul,
	year = {2015},
	pages = {182},
}

@article{paulin-henriksson_point_2008,
	title = {Point spread function calibration requirements for dark energy from cosmic shear},
	volume = {484},
	issn = {0004-6361, 1432-0746},
	url = {http://www.aanda.org/10.1051/0004-6361:20079150},
	doi = {10.1051/0004-6361:20079150},
	number = {1},
	journal = {Astronomy \& Astrophysics},
	author = {Paulin-Henriksson, S. and Amara, A. and Voigt, L. and Refregier, A. and Bridle, S. L.},
	month = jun,
	year = {2008},
	pages = {67--77},
}

@ARTICLE{mdetlsst2023,
       author = {{Sheldon}, Erin S. and {Becker}, Matthew R. and {Jarvis}, Michael and {Armstrong}, Robert and {the LSST Dark Energy Science Collaboration}},
        title = "{Metadetection Weak Lensing for the Vera C. Rubin Observatory}",
      journal = {The Open Journal of Astrophysics},
         year = 2023,
        month = may,
       volume = {6},
          eid = {17},
        pages = {17},
          doi = {10.21105/astro.2303.03947},
archivePrefix = {arXiv},
       eprint = {2303.03947},
 primaryClass = {astro-ph.IM},
       adsurl = {https://ui.adsabs.harvard.edu/abs/2023OJAp....6E..17S}
}

@ARTICLE{Moffat1969,
   author = {{Moffat}, A.~F.~J.},
    title = "{A Theoretical Investigation of Focal Stellar Images in the Photographic Emulsion and Application to Photographic Photometry}",
  journal = {\aap},
     year = 1969,
    month = dec,
   volume = 3,
    pages = {455},
   adsurl = {http://adsabs.harvard.edu/abs/1969A%26A.....3..455M}
}

@article{DESCWLSanchez2021,
    doi = {10.1088/1475-7516/2021/07/043},
    url = {https://doi.org/10.1088%2F1475-7516%2F2021%2F07%2F043},
    year = 2021,
    month = {jul},
    publisher = {{IOP} Publishing},
    volume = {2021},
    number = {07},
    pages = {043},
    author = {Javier Sanchez and Ismael Mendoza and David P. Kirkby and Patricia R. Burchat},
    title = {Effects of overlapping sources on cosmic shear estimation: Statistical sensitivity and pixel-noise bias},
    journal = {Journal of Cosmology and Astroparticle Physics}
}

@ARTICLE{devauc1948,
   author = {{de Vaucouleurs}, G.},
    title = "{Recherches sur les Nebuleuses Extragalactiques}",
  journal = {Annales d'Astrophysique},
     year = 1948,
    month = jan,
   volume = 11,
    pages = {247},
   adsurl = {http://adsabs.harvard.edu/abs/1948AnAp...11..247D}
}

@ARTICLE{SheldonMcal2017,
   author = {{Sheldon}, E.~S. and {Huff}, E.~M.},
    title = "{Practical Weak-lensing Shear Measurement with Metacalibration}",
  journal = {\apj},
archivePrefix = "arXiv",
   eprint = {1702.02601},
     year = 2017,
    month = may,
   volume = 841,
      eid = {24},
    pages = {24},
      doi = {10.3847/1538-4357/aa704b},
   adsurl = {http://adsabs.harvard.edu/abs/2017ApJ...841...24S}
}

@article{hebert_generation_2024,
	title = {Generation of realistic input parameters for simulating atmospheric point-spread functions at astronomical observatories},
	volume = {7},
	copyright = {http://creativecommons.org/licenses/by/4.0},
	issn = {2565-6120},
	url = {https://astro.theoj.org/article/115727-generation-of-realistic-input-parameters-for-simulating-atmospheric-point-spread-functions-at-astronomical-observatories},
	doi = {10.33232/001c.115727},
	journal = {The Open Journal of Astrophysics},
	author = {Hébert, Claire-Alice and Meyers, Joshua E. and Do, My H. and Burchat, Patricia R. and {the LSST Dark Energy Science Collaboration}},
	month = apr,
	year = {2024},
}

@article{schutt_dark_2025,
	author = {Schutt, T. and Jarvis, M. and Roodman, A. and Amon, A. and Becker, M. R. and Gruendl, R. A. and Yamamoto, M. and Bechtol, K. and Bernstein, G. M. and Gatti, M. and Rykoff, E. S. and Sheldon, E. and Troxel, M. A. and Abbott, T. M. C. and Aguena, M. and Alarcon, A. and Andrade-Oliveira, F. and Brooks, D. and Rosell, A. Carnero and Carretero, J. and Chang, C. and Choi, A. and Crocce, M. and da Costa, L. N. and Davis, T. M. and De Vicente, J. and Desai, S. and Diehl, H. T. and Dodelson, S. and Doel, P. and Doux, C. and Drlica-Wagner, A. and Fert{\' e}, A. and Frieman, J. and Garc{\' i}a-Bellido, J. and Gaztanaga, E. and Giannini, G. and Gruen, D. and Gutierrez, G. and Hartley, W.G. and Herner, K. and Hinton, S. R. and Hollowood, D. L. and Honscheid, K. and Huterer, D. and Krause, E. and Kuehn, K. and Lahav, O. and Lee, S. and Lima, M. and Marshall, J. L. and Mena-Fern{\' a}ndez, J. and Miquel, R. and Mohr, J. J. and Muir, J. and Myles, J. and Ogando, R. L. C. and Pieres, A. and Malag{\' o}n, A. A. Plazas and Porredon, A. and Raveri, M. and Rodriguez-Monroy, M. and Samuroff, S. and Sanchez, E. and Cid, D. Sanchez and Sevilla-Noarbe, I. and Smith, M. and Suchyta, E. and Tarle, G. and Vikram, V. and Walker, A. R. and Weaverdyck, N. and Zhang, Y. and {The DES Collaboration}},
	journal = {The Open Journal of Astrophysics},
	doi = {10.33232/001c.132299},
	year = {2025},
	month = {mar 12},
	publisher = {Maynooth Academic Publishing},
	title = {Dark {Energy} {Survey} {Year} 6 {Results}: Point-{Spread} {Function} {Modeling}},
	volume = {8},
}

@misc{yamamoto_dark_2025,
	title = {Dark {Energy} {Survey} {Year} 6 {Results}: {Cell}-based {Coadds} and {Metadetection} {Weak} {Lensing} {Shape} {Catalogue}},
	shorttitle = {Dark {Energy} {Survey} {Year} 6 {Results}},
	url = {http://arxiv.org/abs/2501.05665},
	doi = {10.48550/arXiv.2501.05665},
	publisher = {arXiv},
	author = {Yamamoto, M. and Becker, M. R. and Sheldon, E. and Jarvis, M. and Gruendl, R. A. and Menanteau, F. and Rykoff, E. S. and Mau, S. and Schutt, T. and Gatti, M. and Troxel, M. A. and Amon, A. and Anbajagane, D. and Bernstein, G. M. and Gruen, D. and Huff, E. M. and Tabbutt, M. and Tong, A. and Yanny, B. and Abbott, T. M. C. and Aguena, M. and Alarcon, A. and Andrade-Oliveira, F. and Bechtol, K. and Blazek, J. and Brooks, D. and Rosell, A. Carnero and Carretero, J. and Chang, C. and Choi, A. and Costanzi, M. and Crocce, M. and Costa, L. N. da and Davis, T. M. and Vicente, J. De and Desai, S. and Diehl, H. T. and Dodelson, S. and Doel, P. and Doux, C. and Drlica-Wagner, A. and Ferté, A. and Flaugher, B. and Frieman, J. and García-Bellido, J. and Gaztanaga, E. and Giannini, G. and Gutierrez, G. and Hartley, W. G. and Herner, K. and Hinton, S. R. and Hollowood, D. L. and Honscheid, K. and Huterer, D. and Krause, E. and Kuehn, K. and Lahav, O. and Lima, M. and Marshall, J. L. and Mena-Fernández, J. and Miquel, R. and Mohr, J. J. and Muir, J. and Myles, J. and Ogando, R. L. C. and Pieres, A. and Malagón, A. A. Plazas and Porredon, A. and Prat, J. and Raveri, M. and Rodriguez-Monroy, M. and Roodman, A. and Samuroff, S. and Sanchez, E. and Cid, D. Sanchez and Scarpine, V. and Sevilla-Noarbe, I. and Smith, M. and Suchyta, E. and Tarle, G. and Vikram, V. and Weaverdyck, N. and Wiseman, P. and Zhang, Y.},
	month = jan,
	year = {2025},
	note = {arXiv:2501.05665 [astro-ph]},
}

@ARTICLE{bosch_pipelines_2018,
       author = {{Bosch}, James and {Armstrong}, Robert and {Bickerton}, Steven and {Furusawa}, Hisanori and {Ikeda}, Hiroyuki and {Koike}, Michitaro and {Lupton}, Robert and {Mineo}, Sogo and {Price}, Paul and {Takata}, Tadafumi and {Tanaka}, Masayuki and {Yasuda}, Naoki and {AlSayyad}, Yusra and {Becker}, Andrew C. and {Coulton}, William and {Coupon}, Jean and {Garmilla}, Jose and {Huang}, Song and {Krughoff}, K. Simon and {Lang}, Dustin and {Leauthaud}, Alexie and {Lim}, Kian-Tat and {Lust}, Nate B. and {MacArthur}, Lauren A. and {Mandelbaum}, Rachel and {Miyatake}, Hironao and {Miyazaki}, Satoshi and {Murata}, Ryoma and {More}, Surhud and {Okura}, Yuki and {Owen}, Russell and {Swinbank}, John D. and {Strauss}, Michael A. and {Yamada}, Yoshihiko and {Yamanoi}, Hitomi},
        title = "{The Hyper Suprime-Cam software pipeline}",
      journal = {\pasj},
         year = 2018,
        month = jan,
       volume = {70},
          eid = {S5},
        pages = {S5},
          doi = {10.1093/pasj/psx080},
archivePrefix = {arXiv},
       eprint = {1705.06766},
 primaryClass = {astro-ph.IM},
       adsurl = {https://ui.adsabs.harvard.edu/abs/2018PASJ...70S...5B}
}

@INPROCEEDINGS{bosch_pipelines_2019,
       author = {{Bosch}, James and {AlSayyad}, Yusra and {Armstrong}, Robert and {Bellm}, Eric and {Chiang}, Hsin-Fang and {Eggl}, Siegfried and {Findeisen}, Krzysztof and {Fisher-Levine}, Merlin and {Guy}, Leanne P. and {Guyonnet}, Augustin and {Ivezi{\'c}}, {\v{Z}}eljko and {Jenness}, Tim and {Kov{\'a}cs}, G{\'a}bor and {Krughoff}, K. Simon and {Lupton}, Robert H. and {Lust}, Nate B. and {MacArthur}, Lauren A. and {Meyers}, Joshua and {Moolekamp}, Fred and {Morrison}, Christopher B. and {Morton}, Timothy D. and {O'Mullane}, William and {Parejko}, John K. and {Plazas}, Andr{\'e}s A. and {Price}, Paul A. and {Rawls}, Meredith L. and {Reed}, Sophie L. and {Schellart}, Pim and {Slater}, Colin T. and {Sullivan}, Ian and {Swinbank}, John D. and {Taranu}, Dan and {Waters}, Christopher Z. and {Wood-Vasey}, W.~M.},
        title = "{An Overview of the LSST Image Processing Pipelines}",
    booktitle = {Astronomical Data Analysis Software and Systems XXVII},
         year = 2019,
       series = {Astronomical Society of the Pacific Conference Series},
       volume = {523},
        month = oct,
        pages = {521},
          doi = {10.48550/arXiv.1812.03248},
archivePrefix = {arXiv},
       eprint = {1812.03248},
 primaryClass = {astro-ph.IM},
       adsurl = {https://ui.adsabs.harvard.edu/abs/2019ASPC..523..521B}
}

@article{rowe_improving_2010,
	title = {Improving {PSF} modelling for weak gravitational lensing using new methods in model selection},
	issn = {00358711, 13652966},
	url = {http://mnras.oxfordjournals.org/cgi/doi/10.1111/j.1365-2966.2010.16277.x},
	doi = {10.1111/j.1365-2966.2010.16277.x},
	journal = {Monthly Notices of the Royal Astronomical Society},
	author = {Rowe, Barnaby},
	month = feb,
	year = {2010},
}

@article{hamana_cosmological_2020,
	title = {Cosmological constraints from cosmic shear two-point correlation functions with {HSC} survey first-year data},
	volume = {72},
	copyright = {https://academic.oup.com/journals/pages/open\_access/funder\_policies/chorus/standard\_publication\_model},
	issn = {0004-6264, 2053-051X},
	url = {https://academic.oup.com/pasj/article/doi/10.1093/pasj/psz138/5732415},
	doi = {10.1093/pasj/psz138},
	number = {1},
	journal = {Publications of the Astronomical Society of Japan},
	author = {Hamana, Takashi and Shirasaki, Masato and Miyazaki, Satoshi and Hikage, Chiaki and Oguri, Masamune and More, Surhud and Armstrong, Robert and Leauthaud, Alexie and Mandelbaum, Rachel and Miyatake, Hironao and Nishizawa, Atsushi J and Simet, Melanie and Takada, Masahiro and Aihara, Hiroaki and Bosch, James and Komiyama, Yutaka and Lupton, Robert and Murayama, Hitoshi and Strauss, Michael A and Tanaka, Masayuki},
	month = feb,
	year = {2020},
	pages = {16},
}

@article{von_karman_progress_1948,
	title = {Progress in the {Statistical} {Theory} of {Turbulence}},
	volume = {34},
	issn = {0027-8424, 1091-6490},
	url = {http://www.pnas.org/cgi/doi/10.1073/pnas.34.11.530},
	doi = {10.1073/pnas.34.11.530},
	language = {en},
	number = {11},
	journal = {Proceedings of the National Academy of Sciences},
	author = {von Kármán, T.},
	month = nov,
	year = {1948},
	pages = {530--539},
}

@article{sheldon_mitigating_2020,
	title = {Mitigating {Shear}-dependent {Object} {Detection} {Biases} with {Metacalibration}},
	volume = {902},
	issn = {0004-637X, 1538-4357},
	url = {http://arxiv.org/abs/1911.02505},
	doi = {10.3847/1538-4357/abb595},
	number = {2},
	journal = {The Astrophysical Journal},
	author = {Sheldon, Erin S. and Becker, Matthew R. and MacCrann, Niall and Jarvis, Michael},
	month = oct,
	year = {2020},
	pages = {138},
}

@article{li_differentiable_2023,
	title = {A differentiable perturbation-based weak lensing shear estimator},
	volume = {527},
	copyright = {https://creativecommons.org/licenses/by/4.0/},
	issn = {0035-8711, 1365-2966},
	url = {https://academic.oup.com/mnras/article/527/4/10388/7478002},
	doi = {10.1093/mnras/stad3895},
	number = {4},
	journal = {Monthly Notices of the Royal Astronomical Society},
	author = {Li, Xiangchong and Mandelbaum, Rachel and Jarvis, Mike and Li, Yin and Park, Andy and Zhang, Tianqing},
	month = dec,
	year = {2023},
	pages = {10388--10396},
}

@article{li_analytical_2025,
	title = {Analytical {Noise} {Bias} {Correction} for {Precise} {Weak} {Lensing} {Shear} {Inference}},
	volume = {536},
	issn = {0035-8711, 1365-2966},
	url = {http://arxiv.org/abs/2408.06337},
	doi = {10.1093/mnras/stae2764},
	number = {4},
	journal = {Monthly Notices of the Royal Astronomical Society},
	author = {Li, Xiangchong and Mandelbaum, Rachel and {the LSST Dark Energy Science Collaboration}},
	month = jan,
	year = {2025},
	note = {arXiv:2408.06337 [astro-ph]},
	pages = {3663--3676},
}

@ARTICLE{Zernike1934,
       author = {{Zernike}, von F.},
        title = "{Beugungstheorie des schneidenver-fahrens und seiner verbesserten form, der phasenkontrastmethode}",
      journal = {Physica},
         year = 1934,
        month = may,
       volume = {1},
       number = {7},
        pages = {689-704},
          doi = {10.1016/S0031-8914(34)80259-5},
       adsurl = {https://ui.adsabs.harvard.edu/abs/1934Phy.....1..689Z}
}

@ARTICLE{cosmoSLICS,
       author = {{Harnois-D{\'e}raps}, J. and {Giblin}, B. and {Joachimi}, B.},
        title = "{Cosmic shear covariance matrix in wCDM: Cosmology matters}",
      journal = {\aap},
         year = 2019,
        month = nov,
       volume = {631},
          eid = {A160},
        pages = {A160},
          doi = {10.1051/0004-6361/201935912},
archivePrefix = {arXiv},
       eprint = {1905.06454},
 primaryClass = {astro-ph.CO},
       adsurl = {https://ui.adsabs.harvard.edu/abs/2019A&A...631A.160H}
}

@ARTICLE{OuterRim,
       author = {{Heitmann}, Katrin and {Finkel}, Hal and {Pope}, Adrian and {Morozov}, Vitali and {Frontiere}, Nicholas and {Habib}, Salman and {Rangel}, Esteban and {Uram}, Thomas and {Korytov}, Danila and {Child}, Hillary and {Flender}, Samuel and {Insley}, Joe and {Rizzi}, Silvio},
        title = "{The Outer Rim Simulation: A Path to Many-core Supercomputers}",
      journal = {\apjs},
         year = 2019,
        month = nov,
       volume = {245},
       number = {1},
          eid = {16},
        pages = {16},
          doi = {10.3847/1538-4365/ab4da1},
archivePrefix = {arXiv},
       eprint = {1904.11970},
 primaryClass = {astro-ph.CO},
       adsurl = {https://ui.adsabs.harvard.edu/abs/2019ApJS..245...16H}
}

@ARTICLE{cosmoDC2,
       author = {{Korytov}, Danila and {Hearin}, Andrew and {Kovacs}, Eve and {Larsen}, Patricia and {Rangel}, Esteban and {Hollowed}, Joseph and {Benson}, Andrew J. and {Heitmann}, Katrin and {Mao}, Yao-Yuan and {Bahmanyar}, Anita and {Chang}, Chihway and {Campbell}, Duncan and {DeRose}, Joseph and {Finkel}, Hal and {Frontiere}, Nicholas and {Gawiser}, Eric and {Habib}, Salman and {Joachimi}, Benjamin and {Lanusse}, Fran{\c{c}}ois and {Li}, Nan and {Mandelbaum}, Rachel and {Morrison}, Christopher and {Newman}, Jeffrey A. and {Pope}, Adrian and {Rykoff}, Eli and {Simet}, Melanie and {To}, Chun-Hao and {Vikraman}, Vinu and {Wechsler}, Risa H. and {White}, Martin and {(the LSST Dark Energy Science Collaboration}},
        title = "{CosmoDC2: A Synthetic Sky Catalog for Dark Energy Science with LSST}",
      journal = {\apjs},
         year = 2019,
        month = dec,
       volume = {245},
       number = {2},
          eid = {26},
        pages = {26},
          doi = {10.3847/1538-4365/ab510c},
archivePrefix = {arXiv},
       eprint = {1907.06530},
 primaryClass = {astro-ph.CO},
       adsurl = {https://ui.adsabs.harvard.edu/abs/2019ApJS..245...26K}
}

@ARTICLE{KiDS1000_Joachimi,
       author = {{Joachimi}, B. and {Lin}, C. -A. and {Asgari}, M. and {Tr{\"o}ster}, T. and {Heymans}, C. and {Hildebrandt}, H. and {K{\"o}hlinger}, F. and {S{\'a}nchez}, A.~G. and {Wright}, A.~H. and {Bilicki}, M. and {Blake}, C. and {van den Busch}, J.~L. and {Crocce}, M. and {Dvornik}, A. and {Erben}, T. and {Getman}, F. and {Giblin}, B. and {Hoekstra}, H. and {Kannawadi}, A. and {Kuijken}, K. and {Napolitano}, N.~R. and {Schneider}, P. and {Scoccimarro}, R. and {Sellentin}, E. and {Shan}, H.~Y. and {von Wietersheim-Kramsta}, M. and {Zuntz}, J.},
        title = "{KiDS-1000 methodology: Modelling and inference for joint weak gravitational lensing and spectroscopic galaxy clustering analysis}",
      journal = {\aap},
         year = 2021,
        month = feb,
       volume = {646},
          eid = {A129},
        pages = {A129},
          doi = {10.1051/0004-6361/202038831},
archivePrefix = {arXiv},
       eprint = {2007.01844},
 primaryClass = {astro-ph.CO},
       adsurl = {https://ui.adsabs.harvard.edu/abs/2021A&A...646A.129J}
}

@ARTICLE{LSST_SRD,
       author = {{LSST Dark Energy Science Collaboration} and {Mandelbaum}, Rachel and {Eifler}, Tim and {Hlo{\v{z}}ek}, Ren{\'e}e and {Collett}, Thomas and {Gawiser}, Eric and {Scolnic}, Daniel and {Alonso}, David and {Awan}, Humna and {Biswas}, Rahul and {Blazek}, Jonathan and {Burchat}, Patricia and {Chisari}, Nora Elisa and {Dell'Antonio}, Ian and {Digel}, Seth and {Frieman}, Josh and {Goldstein}, Daniel A. and {Hook}, Isobel and {Ivezi{\'c}}, {\v{Z}}eljko and {Kahn}, Steven M. and {Kamath}, Sowmya and {Kirkby}, David and {Kitching}, Thomas and {Krause}, Elisabeth and {Leget}, Pierre-Fran{\c{c}}ois and {Marshall}, Philip J. and {Meyers}, Joshua and {Miyatake}, Hironao and {Newman}, Jeffrey A. and {Nichol}, Robert and {Rykoff}, Eli and {Sanchez}, F. Javier and {Slosar}, An{\v{z}}e and {Sullivan}, Mark and {Troxel}, M.~A.},
        title = "{The LSST Dark Energy Science Collaboration (DESC) Science Requirements Document}",
      journal = {arXiv e-prints},
         year = 2018,
        month = sep,
          eid = {arXiv:1809.01669},
        pages = {arXiv:1809.01669},
          doi = {10.48550/arXiv.1809.01669},
archivePrefix = {arXiv},
       eprint = {1809.01669},
 primaryClass = {astro-ph.CO},
       adsurl = {https://ui.adsabs.harvard.edu/abs/2018arXiv180901669T}
}

@ARTICLE{Halofit2012,
       author = {{Takahashi}, Ryuichi and {Sato}, Masanori and {Nishimichi}, Takahiro and {Taruya}, Atsushi and {Oguri}, Masamune},
        title = "{Revising the Halofit Model for the Nonlinear Matter Power Spectrum}",
      journal = {\apj},
         year = 2012,
        month = dec,
       volume = {761},
       number = {2},
          eid = {152},
        pages = {152},
          doi = {10.1088/0004-637X/761/2/152},
archivePrefix = {arXiv},
       eprint = {1208.2701},
 primaryClass = {astro-ph.CO},
       adsurl = {https://ui.adsabs.harvard.edu/abs/2012ApJ...761..152T}
}

@ARTICLE{RubinPipe2025,
       author = {{Rubin Observatory Science Pipelines Developers}},
        title = "{The LSST Science Pipelines Software: Optical Survey Pipeline Reduction and Analysis Environment}",
      journal = {Project Science Technical Note PSTN-019, NSF-DOE Vera C. Rubin Observatory},
         year = 2025,
          doi = {10.71929/rubin/2570545},
}

@article{jefferson_reanalysis_2025,
	title = {Reanalysis of {Stage}-{III} cosmic shear surveys: {A} comprehensive study of shear diagnostic tests},
	volume = {8},
	issn = {2565-6120},
	shorttitle = {Reanalysis of {Stage}-{III} cosmic shear surveys},
	url = {http://arxiv.org/abs/2505.03964},
	doi = {10.33232/001c.144668},
	language = {en},
	urldate = {2025-09-29},
	journal = {The Open Journal of Astrophysics},
	author = {Jefferson, Jazmine and Omori, Yuuki and Chang, Chihway and Agarwal, Shrihan and Zuntz, Joe and Asgari, Marika and Gatti, Marco and Giblin, Benjamin and Hébert, Claire-Alice and Jarvis, Mike and Pedersen, Eske M. and Prat, Judit and Schutt, Theo and Zhang, Tianqing and Collaboration, the LSST Dark Energy Science},
	month = sep,
	year = {2025},
}

@inproceedings{connolly_end_2014,
	title = {An end-to-end simulation framework for the {Large} {Synoptic} {Survey} {Telescope}},
	url = {http://proceedings.spiedigitallibrary.org/proceeding.aspx?doi=10.1117/12.2054953},
	doi = {10.1117/12.2054953},
	author = {Connolly, Andrew J. and Angeli, George Z. and Chandrasekharan, Srinivasan and Claver, Charles F. and Cook, Kem and Ivezi\'c, \v{Z}eljko and Jones, R. Lynne and Krughoff, K. Simon and Peng, En-Hsin and Peterson, John and Petry, Catherine and Rasmussen, Andrew P. and Ridgway, Stephen T. and Saha, Abhijit and Sembroski, Glenn and vanderPlas, Jacob and Yoachim, Peter},
    booktitle = {Proc. SPIE 9150},
	month = aug,
	year = {2014},
}

@inproceedings{yoachim_optical_2016,
	title = {An optical to {IR} sky brightness model for the {LSST}},
	url = {http://proceedings.spiedigitallibrary.org/proceeding.aspx?doi=10.1117/12.2232947},
	doi = {10.1117/12.2232947},
	author = {Yoachim, Peter and Coughlin, Michael and Angeli, George Z. and Claver, Charles F. and Connolly, Andrew J. and Cook, Kem and Daniel, Scott and Ivezi\'c, \v{Z}eljko and Jones, R. Lynne and Petry, Catherine and Reuter, Michael and Stubbs, Christopher and Xin, Bo},
	booktitle = {Proc. SPIE 9910},
	month = jul,
	year = {2016},
}

@misc{armstrong_little_2024,
	title = {The little coadd that could: {Estimating} shear from coadded images},
	shorttitle = {The little coadd that could},
	url = {http://arxiv.org/abs/2407.01771},
	publisher = {arXiv},
	author = {Armstrong, Robert and Sheldon, Erin and Huff, Eric and Bosch, Jim and Rykoff, Eli and Mandelbaum, Rachel and Kannawadi, Arun and Melchior, Peter and Lupton, Robert and Becker, Matthew R. and Al-Sayyed, Yusra and Collaboration, the LSST Dark Energy Science},
	month = jul,
	year = {2024},
}

@article{mandelbaum_psfs_2023,
	title = {{PSFs} of coadded images},
	volume = {6},
	issn = {2565-6120},
	url = {http://arxiv.org/abs/2209.09253},
	doi = {10.21105/astro.2209.09253},
	journal = {The Open Journal of Astrophysics},
	author = {Mandelbaum, Rachel and Jarvis, Mike and Lupton, Robert H. and Bosch, James and Kannawadi, Arun and Murphy, Michael D. and Zhang, Tianqing and Collaboration, the LSST Dark Energy Science},
	month = feb,
	year = {2023},
}

@article{bernstein_shapes_2002,
	title = {Shapes and {Shears}, {Stars} and {Smears}: {Optimal} {Measurements} for {Weak} {Lensing}},
	volume = {123},
	issn = {00046256, 15383881},
	shorttitle = {Shapes and {Shears}, {Stars} and {Smears}},
	url = {http://arxiv.org/abs/astro-ph/0107431},
	doi = {10.1086/338085},
	number = {2},
	journal = {The Astronomical Journal},
	author = {Bernstein, G. M. and Jarvis, M.},
	month = feb,
	year = {2002},
	note = {arXiv:astro-ph/0107431},
	pages = {583--618},
}

@article{liaudat_point_2023,
	title = {Point spread function modelling for astronomical telescopes: a review focused on weak gravitational lensing studies},
	volume = {10},
	issn = {2296-987X},
	shorttitle = {Point spread function modelling for astronomical telescopes},
	url = {https://www.frontiersin.org/articles/10.3389/fspas.2023.1158213/full},
	doi = {10.3389/fspas.2023.1158213},
	journal = {Frontiers in Astronomy and Space Sciences},
	author = {Liaudat, Tobías I. and Starck, Jean-Luc and Kilbinger, Martin},
	month = oct,
	year = {2023},
	pages = {1158213},
}

@INPROCEEDINGS{bertin_automated_2011,
       author = {{Bertin}, E.},
        title = "{Automated Morphometry with SExtractor and PSFEx}",
    booktitle = {Astronomical Data Analysis Software and Systems XX},
         year = 2011,
       editor = {{Evans}, I.~N. and {Accomazzi}, A. and {Mink}, D.~J. and {Rots}, A.~H.},
       series = {Astronomical Society of the Pacific Conference Series},
       volume = {442},
        month = jul,
        pages = {435},
       adsurl = {https://ui.adsabs.harvard.edu/abs/2011ASPC..442..435B}
}

@article{miller_bayesian_2013,
	title = {Bayesian galaxy shape measurement for weak lensing surveys – {III}. {Application} to the {Canada}–{France}–{Hawaii} {Telescope} {Lensing} {Survey}},
	volume = {429},
	issn = {1365-2966, 0035-8711},
	url = {http://academic.oup.com/mnras/article/429/4/2858/1008446/Bayesian-galaxy-shape-measurement-for-weak-lensing},
	doi = {10.1093/mnras/sts454},
	number = {4},
	journal = {Monthly Notices of the Royal Astronomical Society},
	author = {Miller, L. and Heymans, C. and Kitching, T. D. and Van Waerbeke, L. and Erben, T. and Hildebrandt, H. and Hoekstra, H. and Mellier, Y. and Rowe, B. T. P. and Coupon, J. and Dietrich, J. P. and Fu, L. and Harnois-Déraps, J. and Hudson, M. J. and Kilbinger, M. and Kuijken, K. and Schrabback, T. and Semboloni, E. and Vafaei, S. and Velander, M.},
	month = mar,
	year = {2013},
}

@article{heymans_impact_2012,
	title = {The impact of high spatial frequency atmospheric distortions on weak lensing measurements},
	issn = {00358711},
	url = {http://arxiv.org/abs/1110.4913},
	doi = {10.1111/j.1365-2966.2011.20312.x},
	journal = {Monthly Notices of the Royal Astronomical Society},
	author = {Heymans, Catherine and Rowe, Barnaby and Hoekstra, Henk and Miller, Lance and Erben, Thomas and Kitching, Thomas and Van Waerbeke, Ludovic},
	month = feb,
	year = {2012},
	note = {arXiv: 1110.4913},
}

@article{massey_origins_2013,
	title = {Origins of weak lensing systematics, and requirements on future instrumentation (or knowledge of instrumentation)},
	volume = {429},
	issn = {1365-2966, 0035-8711},
	url = {http://academic.oup.com/mnras/article/429/1/661/1026608/Origins-of-weak-lensing-systematics-and},
	doi = {10.1093/mnras/sts371},
	journal = {Monthly Notices of the Royal Astronomical Society},
	author = {Massey, Richard and Hoekstra, Henk and Kitching, Thomas and Rhodes, Jason and Cropper, Mark and Amiaux, Jérôme and Harvey, David and Mellier, Yannick and Meneghetti, Massimo and Miller, Lance and Paulin-Henriksson, Stéphane and Pires, Sandrine and Scaramella, Roberto and Schrabback, Tim},
	month = feb,
	year = {2013},
}

@techreport{albrecht_report_2006,
	title = {Report of the {Dark} {Energy} {Task} {Force}},
	url = {http://www.osti.gov/servlets/purl/897600/},
	language = {en},
	author = {Albrecht, Andreas and Bernstein, Gary and Cahn, Robert and Freedman, Wendy L. and Hewitt, Jacqueline and Hu, Wayne and Huth, John and Kamionkowski, Marc and Kolb, Edward W. and Knox, Lloyd and Mather, John C. and Staggs, Suzanne and Suntzeff, Nicholas B.},
	month = sep,
	year = {2006},
	doi = {10.2172/897600},
}

@article{asgari_kids-1000_2021,
	title = {{KiDS}-1000 cosmology: {Cosmic} shear constraints and comparison between two point statistics},
	volume = {645},
	issn = {0004-6361, 1432-0746},
	url = {https://www.aanda.org/10.1051/0004-6361/202039070},
	doi = {10.1051/0004-6361/202039070},
	journal = {Astronomy \& Astrophysics},
	author = {Asgari, Marika and Lin, Chieh-An and Joachimi, Benjamin and Giblin, Benjamin and Heymans, Catherine and Hildebrandt, Hendrik and Kannawadi, Arun and Stölzner, Benjamin and Tröster, Tilman and Van Den Busch, Jan Luca and Wright, Angus H. and Bilicki, Maciej and Blake, Chris and De Jong, Jelte and Dvornik, Andrej and Erben, Thomas and Getman, Fedor and Hoekstra, Henk and Köhlinger, Fabian and Kuijken, Konrad and Miller, Lance and Radovich, Mario and Schneider, Peter and Shan, HuanYuan and Valentijn, Edwin},
	month = jan,
	year = {2021},
}

@misc{wright_kids-legacy_2025,
	title = {{KiDS}-{Legacy}: {Cosmological} constraints from cosmic shear with the complete {Kilo}-{Degree} {Survey}},
	url = {http://arxiv.org/abs/2503.19441},
	doi = {10.48550/arXiv.2503.19441},
	publisher = {arXiv},
	author = {Wright, Angus H. and Stölzner, Benjamin and Asgari, Marika and Bilicki, Maciej and Giblin, Benjamin and Heymans, Catherine and Hildebrandt, Hendrik and Hoekstra, Henk and Joachimi, Benjamin and Kuijken, Konrad and Li, Shun-Sheng and Reischke, Robert and Wietersheim-Kramsta, Maximilian von and Yoon, Mijin and Burger, Pierre and Chisari, Nora Elisa and Jong, Jelte de and Dvornik, Andrej and Georgiou, Christos and Harnois-Déraps, Joachim and Jalan, Priyanka and William, Anjitha John and Joudaki, Shahab and Lesci, Giorgio Francesco and Linke, Laila and Loureiro, Arthur and Mahony, Constance and Maturi, Matteo and Miller, Lance and Moscardini, Lauro and Napolitano, Nicola R. and Porth, Lucas and Radovich, Mario and Schneider, Peter and Tröster, Tilman and Wittje, Anna and Yan, Ziang and Zhang, Yun-Hao},
	month = mar,
	year = {2025},
	note = {arXiv:2503.19441 [astro-ph]},
}

@article{secco_dark_2022,
	title = {Dark {Energy} {Survey} {Year} 3 {Results}: {Cosmology} from {Cosmic} {Shear} and {Robustness} to {Modeling} {Uncertainty}},
	volume = {105},
	issn = {2470-0010, 2470-0029},
	shorttitle = {Dark {Energy} {Survey} {Year} 3 {Results}},
	url = {http://arxiv.org/abs/2105.13544},
	doi = {10.1103/PhysRevD.105.023515},
	journal = {Physical Review D},
	author = {Secco, L. F. and Samuroff, S. and Krause, E. and Jain, B. and Blazek, J. and Raveri, M. and Campos, A. and Amon, A. and Chen, A. and Doux, C. and Choi, A. and Gruen, D. and Bernstein, G. M. and Chang, C. and DeRose, J. and Myles, J. and Ferté, A. and Lemos, P. and Huterer, D. and Prat, J. and Troxel, M. A. and MacCrann, N. and Liddle, A. R. and Kacprzak, T. and Fang, X. and Sánchez, C. and Pandey, S. and Dodelson, S. and Chintalapati, P. and Hoffmann, K. and Alarcon, A. and Alves, O. and Andrade-Oliveira, F. and Baxter, E. J. and Bechtol, K. and Becker, M. R. and Brandao-Souza, A. and Camacho, H. and Rosell, A. Carnero and Kind, M. Carrasco and Cawthon, R. and Cordero, J. P. and Crocce, M. and Davis, C. and Di Valentino, E. and Drlica-Wagner, A. and Eckert, K. and Eifler, T. F. and Elidaiana, M. and Elsner, F. and Elvin-Poole, J. and Everett, S. and Fosalba, P. and Friedrich, O. and Gatti, M. and Giannini, G. and Gruendl, R. A. and Harrison, I. and Hartley, W. G. and Herner, K. and Huang, H. and Huff, E. M. and Jarvis, M. and Jeffrey, N. and Kuropatkin, N. and Leget, P.-F. and Muir, J. and Mccullough, J. and Alsina, A. Navarro and Omori, Y. and Park, Y. and Porredon, A. and Rollins, R. and Roodman, A. and Rosenfeld, R. and Ross, A. J. and Rykoff, E. S. and Sanchez, J. and Sevilla-Noarbe, I. and Sheldon, E. S. and Shin, T. and Tutusaus, I. and Varga, T. N. and Weaverdyck, N. and Wechsler, R. H. and Yanny, B. and Yin, B. and Zhang, Y. and Zuntz, J. and Abbott, T. M. C. and Aguena, M. and Allam, S. and Annis, J. and Bacon, D. and Bertin, E. and Bhargava, S. and Bridle, S. L. and Brooks, D. and Buckley-Geer, E. and Burke, D. L. and Carretero, J. and Costanzi, M. and da Costa, L. N. and De Vicente, J. and Diehl, H. T. and Dietrich, J. P. and Doel, P. and Ferrero, I. and Flaugher, B. and Frieman, J. and García-Bellido, J. and Gaztanaga, E. and Gerdes, D. W. and Giannantonio, T. and Gschwend, J. and Gutierrez, G. and Hinton, S. R. and Hollowood, D. L. and Honscheid, K. and Hoyle, B. and James, D. J. and Jeltema, T. and Kuehn, K. and Lahav, O. and Lima, M. and Lin, H. and Maia, M. A. G. and Marshall, J. L. and Martini, P. and Melchior, P. and Menanteau, F. and Miquel, R. and Mohr, J. J. and Morgan, R. and Ogando, R. L. C. and Palmese, A. and Paz-Chinchón, F. and Petravick, D. and Pieres, A. and Malagón, A. A. Plazas and Rodriguez-Monroy, M. and Romer, A. K. and Sanchez, E. and Scarpine, V. and Schubnell, M. and Scolnic, D. and Serrano, S. and Smith, M. and Soares-Santos, M. and Suchyta, E. and Swanson, M. E. C. and Tarle, G. and Thomas, D. and To, C.},
	month = jan,
	year = {2022},
	note = {arXiv:2105.13544 [astro-ph]},
	pages = {023515},
}

@article{troxel_dark_2018,
	title = {Dark {Energy} {Survey} {Year} 1 results: {Cosmological} constraints from cosmic shear},
	volume = {98},
	issn = {2470-0010, 2470-0029},
	shorttitle = {Dark {Energy} {Survey} {Year} 1 results},
	url = {https://link.aps.org/doi/10.1103/PhysRevD.98.043528},
	doi = {10.1103/PhysRevD.98.043528},
	language = {en},
	number = {4},
	journal = {Physical Review D},
	author = {Troxel, M. A. and MacCrann, N. and Zuntz, J. and Eifler, T. F. and Krause, E. and Dodelson, S. and Gruen, D. and Blazek, J. and Friedrich, O. and Samuroff, S. and Prat, J. and Secco, L. F. and Davis, C. and Ferté, A. and DeRose, J. and Alarcon, A. and Amara, A. and Baxter, E. and Becker, M. R. and Bernstein, G. M. and Bridle, S. L. and Cawthon, R. and Chang, C. and Choi, A. and De Vicente, J. and Drlica-Wagner, A. and Elvin-Poole, J. and Frieman, J. and Gatti, M. and Hartley, W. G. and Honscheid, K. and Hoyle, B. and Huff, E. M. and Huterer, D. and Jain, B. and Jarvis, M. and Kacprzak, T. and Kirk, D. and Kokron, N. and Krawiec, C. and Lahav, O. and Liddle, A. R. and Peacock, J. and Rau, M. M. and Refregier, A. and Rollins, R. P. and Rozo, E. and Rykoff, E. S. and Sánchez, C. and Sevilla-Noarbe, I. and Sheldon, E. and Stebbins, A. and Varga, T. N. and Vielzeuf, P. and Wang, M. and Wechsler, R. H. and Yanny, B. and Abbott, T. M. C. and Abdalla, F. B. and Allam, S. and Annis, J. and Bechtol, K. and Benoit-Lévy, A. and Bertin, E. and Brooks, D. and Buckley-Geer, E. and Burke, D. L. and Carnero Rosell, A. and Carrasco Kind, M. and Carretero, J. and Castander, F. J. and Crocce, M. and Cunha, C. E. and D’Andrea, C. B. and Da Costa, L. N. and DePoy, D. L. and Desai, S. and Diehl, H. T. and Dietrich, J. P. and Doel, P. and Fernandez, E. and Flaugher, B. and Fosalba, P. and García-Bellido, J. and Gaztanaga, E. and Gerdes, D. W. and Giannantonio, T. and Goldstein, D. A. and Gruendl, R. A. and Gschwend, J. and Gutierrez, G. and James, D. J. and Jeltema, T. and Johnson, M. W. G. and Johnson, M. D. and Kent, S. and Kuehn, K. and Kuhlmann, S. and Kuropatkin, N. and Li, T. S. and Lima, M. and Lin, H. and Maia, M. A. G. and March, M. and Marshall, J. L. and Martini, P. and Melchior, P. and Menanteau, F. and Miquel, R. and Mohr, J. J. and Neilsen, E. and Nichol, R. C. and Nord, B. and Petravick, D. and Plazas, A. A. and Romer, A. K. and Roodman, A. and Sako, M. and Sanchez, E. and Scarpine, V. and Schindler, R. and Schubnell, M. and Smith, M. and Smith, R. C. and Soares-Santos, M. and Sobreira, F. and Suchyta, E. and Swanson, M. E. C. and Tarle, G. and Thomas, D. and Tucker, D. L. and Vikram, V. and Walker, A. R. and Weller, J. and Zhang, Y. and {DES Collaboration}},
	month = aug,
	year = {2018},
	pages = {043528},
}

@article{amon_dark_2022,
	title = {Dark {Energy} {Survey} {Year} 3 {Results}: {Cosmology} from {Cosmic} {Shear} and {Robustness} to {Data} {Calibration}},
	volume = {105},
	issn = {2470-0010, 2470-0029},
	shorttitle = {Dark {Energy} {Survey} {Year} 3 {Results}},
	url = {http://arxiv.org/abs/2105.13543},
	doi = {10.1103/PhysRevD.105.023514},
	journal = {Physical Review D},
	author = {Amon, A. and Gruen, D. and Troxel, M. A. and MacCrann, N. and Dodelson, S. and Choi, A. and Doux, C. and Secco, L. F. and Samuroff, S. and Krause, E. and Cordero, J. and Myles, J. and DeRose, J. and Wechsler, R. H. and Gatti, M. and Navarro-Alsina, A. and Bernstein, G. M. and Jain, B. and Blazek, J. and Alarcon, A. and Ferté, A. and Raveri, M. and Lemos, P. and Campos, A. and Prat, J. and Sánchez, C. and Jarvis, M. and Alves, O. and Andrade-Oliveira, F. and Baxter, E. and Bechtol, K. and Becker, M. R. and Bridle, S. L. and Camacho, H. and Campos, A. and Rosell, A. Carnero and Kind, M. Carrasco and Cawthon, R. and Chang, C. and Chen, R. and Chintalapati, P. and Crocce, M. and Davis, C. and Diehl, H. T. and Drlica-Wagner, A. and Eckert, K. and Eifler, T. F. and Elvin-Poole, J. and Everett, S. and Fang, X. and Fosalba, P. and Friedrich, O. and Giannini, G. and Gruendl, R. A. and Harrison, I. and Hartley, W. G. and Herner, K. and Huang, H. and Huff, E. M. and Huterer, D. and Kuropatkin, N. and Leget, P.-F. and Liddle, A. R. and McCullough, J. and Muir, J. and Pandey, S. and Park, Y. and Porredon, A. and Refregier, A. and Rollins, R. P. and Roodman, A. and Rosenfeld, R. and Ross, A. J. and Rykoff, E. S. and Sanchez, J. and Sevilla-Noarbe, I. and Sheldon, E. and Shin, T. and Troja, A. and Tutusaus, I. and Varga, T. N. and Weaverdyck, N. and Yanny, B. and Yin, B. and Zhang, Y. and Zuntz, J. and Aguena, M. and Allam, S. and Annis, J. and Bacon, D. and Bertin, E. and Bhargava, S. and Brooks, D. and Buckley-Geer, E. and Burke, D. L. and Carretero, J. and Costanzi, M. and da Costa, L. N. and Pereira, M. E. S. and De Vicente, J. and Desai, S. and Dietrich, J. P. and Doel, P. and Ferrero, I. and Flaugher, B. and Frieman, J. and García-Bellido, J. and Gaztanaga, E. and Gerdes, D. W. and Giannantonio, T. and Gschwend, J. and Gutierrez, G. and Hinton, S. R. and Hollowood, D. L. and Honscheid, K. and Hoyle, B. and James, D. J. and Kron, R. and Kuehn, K. and Lahav, O. and Lima, M. and Lin, H. and Maia, M. A. G. and Marshall, J. L. and Martini, P. and Melchior, P. and Menanteau, F. and Miquel, R. and Mohr, J. J. and Morgan, R. and Ogando, R. L. C. and Palmese, A. and Paz-Chinchón, F. and Petravick, D. and Pieres, A. and Malagón, A. A. Plazas and Romer, A. K. and Sanchez, E. and Scarpine, V. and Schubnell, M. and Serrano, S. and Smith, M. and Soares-Santos, M. and Suchyta, E. and Tarle, G. and Thomas, D. and To, C. and Weller, J.},
	month = jan,
	year = {2022},
}

@article{kilbinger_cosmology_2015,
	title = {Cosmology with cosmic shear observations: a review},
	volume = {78},
	issn = {0034-4885, 1361-6633},
	shorttitle = {Cosmology with cosmic shear observations},
	url = {https://iopscience.iop.org/article/10.1088/0034-4885/78/8/086901},
	doi = {10.1088/0034-4885/78/8/086901},
	journal = {Reports on Progress in Physics},
	author = {Kilbinger, Martin},
	month = jul,
	year = {2015},
	pages = {086901},
}

@article{mandelbaum_weak_2018,
	title = {Weak {Lensing} for {Precision} {Cosmology}},
	volume = {56},
	issn = {0066-4146, 1545-4282},
	url = {https://www.annualreviews.org/doi/10.1146/annurev-astro-081817-051928},
	doi = {10.1146/annurev-astro-081817-051928},
	urldate = {2021-01-07},
	journal = {Annual Review of Astronomy and Astrophysics},
	author = {Mandelbaum, Rachel},
	month = sep,
	year = {2018},
	pages = {393--433},
}

@misc{li_hyper_2023,
	title = {Hyper {Suprime}-{Cam} {Year} 3 {Results}: {Cosmology} from {Cosmic} {Shear} {Two}-point {Correlation} {Functions}},
	shorttitle = {Hyper {Suprime}-{Cam} {Year} 3 {Results}},
	url = {http://arxiv.org/abs/2304.00702},
	publisher = {arXiv},
	author = {Li, Xiangchong and Zhang, Tianqing and Sugiyama, Sunao and Dalal, Roohi and Rau, Markus M. and Mandelbaum, Rachel and Takada, Masahiro and More, Surhud and Strauss, Michael A. and Miyatake, Hironao and Shirasaki, Masato and Hamana, Takashi and Oguri, Masamune and Luo, Wentao and Nishizawa, Atsushi J. and Takahashi, Ryuichi and Nicola, Andrina and Osato, Ken and Kannawadi, Arun and Sunayama, Tomomi and Armstrong, Robert and Komiyama, Yutaka and Lupton, Robert H. and Lust, Nate B. and Miyazaki, Satoshi and Murayama, Hitoshi and Nishimichi, Takahiro and Okura, Yuki and Price, Paul A. and Tait, Philip J. and Tanaka, Masayuki and Wang, Shiang-Yu},
	month = apr,
	year = {2023},
}

@article{ivezic_lsst_2019,
	title = {{LSST}: {From} {Science} {Drivers} to {Reference} {Design} and {Anticipated} {Data} {Products}},
	volume = {873},
	url = {https://doi.org/10.3847/1538-4357/ab042c},
	doi = {10.3847/1538-4357/ab042c},
	number = {2},
	journal = {The Astrophysical Journal},
	author = {Ivezić, \v{Z}eljko and Kahn, Steven M. and Tyson, J. Anthony and Abel, Bob and Acosta, Emily and Allsman, Robyn and Alonso, David and AlSayyad, Yusra and Anderson, Scott F. and Andrew, John and P. Angel, James Roger and Angeli, George Z. and Ansari, Reza and Antilogus, Pierre and Araujo, Constanza and Armstrong, Robert and Arndt, Kirk T. and Astier, Pierre and Aubourg, Éric and Auza, Nicole and Axelrod, Tim S. and Bard, Deborah J. and Barr, Jeff D. and Barrau, Aurelian and Bartlett, James G. and Bauer, Amanda E. and Bauman, Brian J. and Baumont, Sylvain and Bechtol, Ellen and Bechtol, Keith and Becker, Andrew C. and Becla, Jacek and Beldica, Cristina and Bellavia, Steve and Bianco, Federica B. and Biswas, Rahul and Blanc, Guillaume and Blazek, Jonathan and Blandford, Roger D. and Bloom, Josh S. and Bogart, Joanne and Bond, Tim W. and Booth, Michael T. and Borgland, Anders W. and Borne, Kirk and Bosch, James F. and Boutigny, Dominique and Brackett, Craig A. and Bradshaw, Andrew and Brandt, William Nielsen and Brown, Michael E. and Bullock, James S. and Burchat, Patricia and Burke, David L. and Cagnoli, Gianpietro and Calabrese, Daniel and Callahan, Shawn and Callen, Alice L. and Carlin, Jeffrey L. and Carlson, Erin L. and Chandrasekharan, Srinivasan and Charles-Emerson, Glenaver and Chesley, Steve and Cheu, Elliott C. and Chiang, Hsin-Fang and Chiang, James and Chirino, Carol and Chow, Derek and Ciardi, David R. and Claver, Charles F. and Cohen-Tanugi, Johann and Cockrum, Joseph J. and Coles, Rebecca and Connolly, Andrew J. and Cook, Kem H. and Cooray, Asantha and Covey, Kevin R. and Cribbs, Chris and Cui, Wei and Cutri, Roc and Daly, Philip N. and Daniel, Scott F. and Daruich, Felipe and Daubard, Guillaume and Daues, Greg and Dawson, William and Delgado, Francisco and Dellapenna, Alfred and Peyster, Robert de and Val-Borro, Miguel de and Digel, Seth W. and Doherty, Peter and Dubois, Richard and Dubois-Felsmann, Gregory P. and Durech, Josef and Economou, Frossie and Eifler, Tim and Eracleous, Michael and Emmons, Benjamin L. and Neto, Angelo Fausti and Ferguson, Henry and Figueroa, Enrique and Fisher-Levine, Merlin and Focke, Warren and Foss, Michael D. and Frank, James and Freemon, Michael D. and Gangler, Emmanuel and Gawiser, Eric and Geary, John C. and Gee, Perry and Geha, Marla and Gessner, Charles J. B. and Gibson, Robert R. and Gilmore, D. Kirk and Glanzman, Thomas and Glick, William and Goldina, Tatiana and Goldstein, Daniel A. and Goodenow, Iain and Graham, Melissa L. and Gressler, William J. and Gris, Philippe and Guy, Leanne P. and Guyonnet, Augustin and Haller, Gunther and Harris, Ron and Hascall, Patrick A. and Haupt, Justine and Hernandez, Fabio and Herrmann, Sven and Hileman, Edward and Hoblitt, Joshua and Hodgson, John A. and Hogan, Craig and Howard, James D. and Huang, Dajun and Huffer, Michael E. and Ingraham, Patrick and Innes, Walter R. and Jacoby, Suzanne H. and Jain, Bhuvnesh and Jammes, Fabrice and Jee, M. James and Jenness, Tim and Jernigan, Garrett and Jevremović, Darko and Johns, Kenneth and Johnson, Anthony S. and Johnson, Margaret W. G. and Jones, R. Lynne and Juramy-Gilles, Claire and Jurić, Mario and Kalirai, Jason S. and Kallivayalil, Nitya J. and Kalmbach, Bryce and Kantor, Jeffrey P. and Karst, Pierre and Kasliwal, Mansi M. and Kelly, Heather and Kessler, Richard and Kinnison, Veronica and Kirkby, David and Knox, Lloyd and Kotov, Ivan V. and Krabbendam, Victor L. and Krughoff, K. Simon and Kubánek, Petr and Kuczewski, John and Kulkarni, Shri and Ku, John and Kurita, Nadine R. and Lage, Craig S. and Lambert, Ron and Lange, Travis and Langton, J. Brian and Guillou, Laurent Le and Levine, Deborah and Liang, Ming and Lim, Kian-Tat and Lintott, Chris J. and Long, Kevin E. and Lopez, Margaux and Lotz, Paul J. and Lupton, Robert H. and Lust, Nate B. and MacArthur, Lauren A. and Mahabal, Ashish and Mandelbaum, Rachel and Markiewicz, Thomas W. and Marsh, Darren S. and Marshall, Philip J. and Marshall, Stuart and May, Morgan and McKercher, Robert and McQueen, Michelle and Meyers, Joshua and Migliore, Myriam and Miller, Michelle and Mills, David J. and Miraval, Connor and Moeyens, Joachim and Moolekamp, Fred E. and Monet, David G. and Moniez, Marc and Monkewitz, Serge and Montgomery, Christopher and Morrison, Christopher B. and Mueller, Fritz and Muller, Gary P. and Arancibia, Freddy Muñoz and Neill, Douglas R. and Newbry, Scott P. and Nief, Jean-Yves and Nomerotski, Andrei and Nordby, Martin and O’Connor, Paul and Oliver, John and Olivier, Scot S. and Olsen, Knut and O’Mullane, William and Ortiz, Sandra and Osier, Shawn and Owen, Russell E. and Pain, Reynald and Palecek, Paul E. and Parejko, John K. and Parsons, James B. and Pease, Nathan M. and Peterson, J. Matt and Peterson, John R. and Petravick, Donald L. and Petrick, M. E. Libby and Petry, Cathy E. and Pierfederici, Francesco and Pietrowicz, Stephen and Pike, Rob and Pinto, Philip A. and Plante, Raymond and Plate, Stephen and Plutchak, Joel P. and Price, Paul A. and Prouza, Michael and Radeka, Veljko and Rajagopal, Jayadev and Rasmussen, Andrew P. and Regnault, Nicolas and Reil, Kevin A. and Reiss, David J. and Reuter, Michael A. and Ridgway, Stephen T. and Riot, Vincent J. and Ritz, Steve and Robinson, Sean and Roby, William and Roodman, Aaron and Rosing, Wayne and Roucelle, Cecille and Rumore, Matthew R. and Russo, Stefano and Saha, Abhijit and Sassolas, Benoit and Schalk, Terry L. and Schellart, Pim and Schindler, Rafe H. and Schmidt, Samuel and Schneider, Donald P. and Schneider, Michael D. and Schoening, William and Schumacher, German and Schwamb, Megan E. and Sebag, Jacques and Selvy, Brian and Sembroski, Glenn H. and Seppala, Lynn G. and Serio, Andrew and Serrano, Eduardo and Shaw, Richard A. and Shipsey, Ian and Sick, Jonathan and Silvestri, Nicole and Slater, Colin T. and Smith, J. Allyn and Smith, R. Chris and Sobhani, Shahram and Soldahl, Christine and Storrie-Lombardi, Lisa and Stover, Edward and Strauss, Michael A. and Street, Rachel A. and Stubbs, Christopher W. and Sullivan, Ian S. and Sweeney, Donald and Swinbank, John D. and Szalay, Alexander and Takacs, Peter and Tether, Stephen A. and Thaler, Jon J. and Thayer, John Gregg and Thomas, Sandrine and Thornton, Adam J. and Thukral, Vaikunth and Tice, Jeffrey and Trilling, David E. and Turri, Max and Berg, Richard Van and Berk, Daniel Vanden and Vetter, Kurt and Virieux, Francoise and Vucina, Tomislav and Wahl, William and Walkowicz, Lucianne and Walsh, Brian and Walter, Christopher W. and Wang, Daniel L. and Wang, Shin-Yawn and Warner, Michael and Wiecha, Oliver and Willman, Beth and Winters, Scott E. and Wittman, David and Wolff, Sidney C. and Wood-Vasey, W. Michael and Wu, Xiuqin and Xin, Bo and Yoachim, Peter and Zhan, Hu},
	month = mar,
	year = {2019},
	note = {Publisher: The American Astronomical Society},
	pages = {111},
}

@article{esteves_photometry_2023,
	title = {Photometry, {Centroid} and {Point}-spread {Function} {Measurements} in the {LSST} {Camera} {Focal} {Plane} {Using} {Artificial} {Stars}},
	volume = {135},
	issn = {1538-3873},
	url = {https://doi.org/10.1088/1538-3873/ad0a73},
	doi = {10.1088/1538-3873/ad0a73},
	journal = {Publications of the Astronomical Society of the Pacific},
	author = {Esteves, Johnny H. and Utsumi, Yousuke and Snyder, Adam and Schutt, Theo and Broughton, Alex and Trbalic, Bahrudin and Mau, Sidney and Rasmussen, Andrew and Plazas Malagón, Andrés A. and Bradshaw, Andrew and Marshall, Stuart and Digel, Seth and Chiang, James and Rykoff, Eli and Waters, Chris and Soares-Santos, Marcelle and Roodman, Aaron},
	month = dec,
	year = {2023},
	note = {Publisher: The Astronomical Society of the Pacific},
}

@article{reischke_kids_2025,
   title={KiDS-Legacy: Covariance validation and the unified OneCovariance framework for projected large-scale structure observables},
   volume={699},
   ISSN={1432-0746},
   url={http://dx.doi.org/10.1051/0004-6361/202452592},
   DOI={10.1051/0004-6361/202452592},
   journal={Astronomy \&; Astrophysics},
   publisher={EDP Sciences},
   author={Reischke, Robert and Unruh, Sandra and Asgari, Marika and Dvornik, Andrej and Hildebrandt, Hendrik and Joachimi, Benjamin and Porth, Lucas and von Wietersheim-Kramsta, Maximilian and van den Busch, Jan Luca and Stölzner, Benjamin and Wright, Angus H. and Yan, Ziang and Bilicki, Maciej and Burger, Pierre and Chisari, Nora Elisa and Harnois-Déraps, Joachim and Georgiou, Christos and Heymans, Catherine and Jalan, Priyanka and Joudaki, Shahab and Kuijken, Konrad and Li, Shun-Sheng and Linke, Laila and Mahony, Constance and Sciotti, Davide and Tröster, Tilman and Yoon, Mijin},
   year={2025},
   month=jul, pages={A124} }

@article{hirata_ggl_2004,
	title = {Galaxy-galaxy weak lensing in the {Sloan} {Digital} {Sky} {Survey}: intrinsic alignments and shear calibration errors},
	volume = {353},
	issn = {00358711, 13652966},
	url = {https://academic.oup.com/mnras/article-lookup/doi/10.1111/j.1365-2966.2004.08090.x},
	doi = {10.1111/j.1365-2966.2004.08090.x},
	number = {2},
	journal = {Monthly Notices of the Royal Astronomical Society},
	author = {Hirata, Christopher M. and Mandelbaum, Rachel and Seljak, Uroš and Guzik, Jacek and Padmanabhan, Nikhil and Blake, Cullen and Brinkmann, Jonathan and Budávari, Tamas and Connolly, Andrew and Csabai, Istvan and Scranton, Ryan and Szalay, Alexander S.},
	month = sep,
	year = {2004},
	pages = {529--549},
}

@techreport{rubinobs_dp1_2025,
	type = {Rubin {Technical} {Note}},
	title = {The {Vera} {C}. {Rubin} {Observatory} {Data} {Preview} 1},
	url = {https://rtn-095.lsst.io/},
	doi = {10.71929/rubin/2570536},
	number = {RTN-095},
	institution = {NSF-DOE Vera C. Rubin Observatory},
	author = {{NSF-DOE Vera C. Rubin Observatory}},
	month = jul,
	year = {2025},
}
\bibliographystyle{aasjournal}

\end{document}